\documentclass[12pt]{article}
\usepackage{amsmath,amssymb}
\usepackage{bm}
\usepackage{color}
\usepackage{cite}

\def\slash#1{\not\!\!#1}

\begin{document}

\title{
\begin{flushright}
\ \\*[-80pt]
\begin{minipage}{0.2\linewidth}
\normalsize
EPHOU-20-005
\end{minipage}
\end{flushright}
{\Large \bf
Revisiting 
modular symmetry in magnetized \\ torus and orbifold compactifications
\\*[20pt]}}

\author{
Shota Kikuchi,
~Tatsuo Kobayashi,
~Shintaro Takada, \\
~Takuya H. Tatsuishi,
 and~Hikaru Uchida
\\*[20pt]
\centerline{
\begin{minipage}{\linewidth}
\begin{center}
{\it \normalsize
Department of Physics, Hokkaido University, Sapporo 060-0810, Japan} \\*[5pt]
\end{center}
\end{minipage}}
\\*[50pt]}

\date{
\centerline{\small \bf Abstract}
\begin{minipage}{0.9\linewidth}
\medskip
\medskip
\small
We study the modular symmetry in $T^2$ and orbifold comfactifications with 
magnetic fluxes.
There are $|M|$ zero-modes on $T^2$ with the magnetic flux $M$.
Their wavefunctions as well as massive modes behave as modular forms of weight $1/2$ and 
represent the double covering group of $\Gamma \equiv SL(2,\mathbb{Z})$, 
$\widetilde{\Gamma} \equiv \widetilde{SL}(2,\mathbb{Z})$.
Each wavefunction on $T^2$ with the magnetic flux $M$ transforms under
 $\widetilde{\Gamma}(2|M|)$, which is the normal subgroup of $\widetilde{SL}(2,\mathbb{Z})$.
Then, $|M|$ zero-modes are representations of  the quotient group $\widetilde{\Gamma}'_{2|M|} \equiv \widetilde{\Gamma}/\widetilde{\Gamma}(2|M|)$.
We also study the modular symmetry on twisted and shifted orbifolds $T^2/\mathbb{Z}_N$.
Wavefunctions are decomposed into smaller representations by eigenvalues of twist and shift.
They provide us with reduction of reducible representations on $T^2$.
\end{minipage}
}

\begin{titlepage}
\maketitle
\thispagestyle{empty}
\end{titlepage}

\newpage


\section{Introduction}
\label{Intro}

The standard model (SM) is now well established.
However, the origin of the flavor structure of quarks and leptons is still one of the mysteries of the SM.
Various studies have been carried out to understand the flavor structure.
One of the interesting approaches is to impose some non-Abelian discrete flavor symmetries~\cite{
	Altarelli:2010gt,Ishimori:2010au,Ishimori:2012zz,Hernandez:2012ra,
	King:2013eh,King:2014nza,Tanimoto:2015nfa,King:2017guk,Petcov:2017ggy}
on the flavors of quarks and leptons.
Various discrete symmetries such as $S_N$, $A_N$, $\Delta(3N^2)$, $\Delta(6N^2)$ are used.
Then, 
these flavor symmetries are broken by vacuum expectation values (VEVs) of gauge singlet scalars, the so-called flavons in order to realize the masses and the mixing angles of quarks and leptons.
However, a complicated vacuum alignment is required.

Superstring theory predicts six-dimensional (6D) compact space in addition to 
our four-dimensional (4D) space-time.
Such a compact space may provide us with origins of non-Abelian discrete flavor 
symmetries. (See, e.g. \cite{Kobayashi:2006wq,Abe:2009vi}.)
In particular, the torus as well as orbifolds has the modular symmetry 
as geometrical symmetry.
Zero-modes transform under the modular symmetry.
That is, the modular symmetry is a flavor symmetry in a sense.
This transformation behavior has been studied in magnetized D-brane models \cite{Kobayashi:2017dyu,Kobayashi:2018rad,Kobayashi:2018bff,Ohki:2020bpo}
and heterotic orbifold models \cite{Lauer:1989ax,Lerche:1989cs,Ferrara:1989qb,Baur:2019kwi,Nilles:2020nnc}.
(See also \cite{Kobayashi:2016ovu,Kariyazono:2019ehj,Kobayashi:2020hoc}.)
However, the modular flavor symmetry is different from 
the conventional flavor symmetries.
Yukawa couplings as well as higher order couplings are not singlets, but transform 
under the modular symmetry.

Interestingly, the modular symmetry includes the finite modular groups $\Gamma_N$ for $N=2,3,4,5$, which are isomorphic to $S_3$, $A_4$, $S_4$, $A_5$~\cite{deAdelhartToorop:2011re}, respectively.
Recently inspired by these aspects, a new bottom-up approach of flavor models has been 
studied extensively \cite{Feruglio:2017spp,Kobayashi:2018vbk,Penedo:2018nmg,Kobayashi:2018scp,deAnda:2018ecu,Okada:2018yrn,Ding:2019xna,
Nomura:2019jxj,Novichkov:2019sqv,Liu:2019khw,Asaka:2019vev,Zhang:2019ngf,Wang:2019ovr,Kobayashi:2019gtp,Lu:2019vgm,Wang:2019xbo,Abbas:2020qzc}.
In those models, some finite modular groups are applied as the flavor symmetries.
Furthermore, it is notable that the Yukawa couplings are functions of the modulus $\tau$, 
which describes the shape of the compact space,  
and are assigned to 
modular forms, which transform non-trivially under the modular transformations.
The flavor modular symmetry can be  broken by the VEV of the modulus $\tau$ without flavons.

As mentioned above, the modular symmetry is quite important from both top-down and bottom-up 
approaches.
That could become a bridge to connect high and low energy scales.
Our purpose of this paper is to study the modular symmetry in more detail.
We study how wavefunctions on $T^2$ with magnetic flux transform under the modular 
symmetry.
Furthermore, we also study twisted and shifted orbifolds.
Orbifold twist and shift decompose wavefunctions by their eigenvalues.
That provides us with reduction of reducible representations.
Also, it provides us with a new approach to construct three-generation models from 
the phenomenological viewpoint. 

This paper is organized as follows.
In section~\ref{modularsym}, we briefly review the modular symmetry on $T^2$ and modular forms.
After reviewing $T^2$ with magnetic flux in section~\ref{wavT2}, we study the modular symmetry on the magnetized $T^2$ in section~\ref{modularT2}.
We find that the wavefunctions on the magnetized $T^2$ are transformed under the modular transformations like modular forms of weight $1/2$ for $\widetilde{\Gamma}(2|M|)$.
We also study the modular symmetry on various magnetized $T^2/\mathbb{Z}_N$ orbifolds in sections~\ref{T2ZNtwist}-\ref{T2Z2twistshift}.
In section~\ref{T2ZNtwist}, we study that on the $T^2/\mathbb{Z}_N$ twisted orbifolds.
In section~\ref{T2ZNshift}, we study that on the $T^2/\mathbb{Z}_N$ shifted orbifolds.
Furthermore, in section~\ref{T2Z2twistshift}, we study that on the $T^2/\mathbb{Z}_N$ twisted and shifted orbifolds.
In those sections, we find that the modular symmetry remains on the $T^2/\mathbb{Z}_2$ twisted orbifold and the ''full'' shifted $T^2/\mathbb{Z}_N$ orbifolds.
In particular, the full $T^2/\mathbb{Z}_2$ shifted orbifold is consistent with the $T^2/\mathbb{Z}_2$ twisted orbifold.
Section~\ref{conclusion} is the conclusion.
Appendix~\ref{gCP} shows the extension for the generalized $CP$ symmetry with the modular symmetry on the magnetized $T^2$.
Appendix~\ref{withgauge} shows the detail calculation of discussion in section~\ref{modularT2}.
Appendix~\ref{exM816} shows examples of the magnetized $T^2/\mathbb{Z}_2$ twisted and shifted orbifold models.


\section{Modular symmetry and modular forms}
\label{modularsym}

In this section, we briefly review the modular symmetry on $T^2$ and the modular forms.
(See e.g \cite{Gunning:1962,Schoeneberg:1974,Koblitz:1984,Bruinier:2008}. See also \cite{Liu:2019khw}.)
First, we review the modular symmetry on $T^2$.
The torus $T^2$ can be constructed as division of the complex plane $\mathbb{C}$ 
by a two-dimensional (2D) lattice $\Lambda$,  i.e.~$T^2 \simeq \mathbb{C}/\Lambda$.
The lattice $\Lambda$ is spanned by two lattice vectors $e_i\ (i=1,2)$.
We denote the complex coordinate of $\mathbb{C}$ as $u$ and that of the $T^2$ as $z \equiv u/e_1$.
We also introduce the complex modulus parameter $\tau \equiv e_2/e_1\ ({\rm Im}\tau > 0)$.
However, there is some ambiguity in choice of the lattice vectors.
The lattice spanned by the following lattice vectors $e'_i\ (i=1,2)$, 
\begin{align}
\begin{pmatrix}
e'_2 \\ e'_1
\end{pmatrix}
=
\begin{pmatrix}
a & b \\
c & d
\end{pmatrix}
\begin{pmatrix}
e_2 \\ e_1
\end{pmatrix},
\quad
\gamma =
\begin{pmatrix}
a & b \\
c & d
\end{pmatrix}
\in SL(2,\mathbb{Z}) \equiv \Gamma,
\label{SL2Z}
\end{align}
is the same lattice spanned by the lattice vectors $e_i\ (i=1,2)$.
Under the above $SL(2,\mathbb{Z})$ transformation, the coordinate of $T^2$ and the modulus are transformed as
\begin{align}
\gamma: z \equiv \frac{u}{e_1} \rightarrow z' \equiv \frac{u}{e'_1} = \frac{z}{c\tau+d}, \label{zSL2Z} \\
\gamma: \tau \equiv \frac{e_2}{e_1} \rightarrow \tau' \equiv \frac{e'_2}{e'_1} = \frac{a\tau+b}{c\tau+d}. \label{tauSL2Z}
\end{align}
This is the modular transformation.
The group $\Gamma \equiv SL(2,\mathbb{Z})$ is generated by two generators,
\begin{align}
S=
\begin{pmatrix}
0 & 1 \\
-1 & 0
\end{pmatrix},
\quad
T=
\begin{pmatrix}
1 & 1 \\
0 & 1
\end{pmatrix}.
\label{SandT}
\end{align}
They satisfy the following algebraic relations\footnote{They satisfy $(ST^{-1})^3=-\mathbb{I}, (ST^{-1})^{6}=\mathbb{I}$.},
\begin{align}
S^2=-\mathbb{I}, \quad S^4=(ST)^3=\mathbb{I}. \label{algebrarel}
\end{align}
Under $S$ and $T$, the coordinate of $T^2$ and the modulus, $(z,\tau)$, are transformed as
\begin{align}
S: (z,\tau) \rightarrow \left( -\frac{z}{\tau}, -\frac{1}{\tau} \right), \quad
T: (z,\tau) \rightarrow (z,\tau+1). \label{ztauSandT}
\end{align}
Note that $-\mathbb{I}:(z,\tau)\rightarrow(-z,\tau)$.

Next, we review the modular forms.
The principal congruence subgroup of level $N$, $\Gamma(N)$ is the normal subgroup of $\Gamma$ defined by
\begin{align}
\Gamma(N) \equiv \left\{ h =
\begin{pmatrix}
a' & b' \\
c' & d'
\end{pmatrix} \in \Gamma \biggl|
\begin{pmatrix}
a' & b' \\
c' & d'
\end{pmatrix}
\equiv
\begin{pmatrix}
1 & 0 \\
0 & 1
\end{pmatrix}
\ ({\rm mod}\ N) \right\}, \label{GammaN}
\end{align}
where we have $\Gamma(1) \simeq \Gamma$.
The modular forms $f(\tau)$ of weight $k$ for $\Gamma(N)$ are holomorphic functions of $\tau$ which transform as
\begin{align}
f(h(\tau)) = (c'\tau+d')^k f(\tau), \quad
h =
\begin{pmatrix}
a' & b' \\
c' & d'
\end{pmatrix}
\in \Gamma(N). \label{modularformGammaN}
\end{align}
Here, $k$ is an integer while $k$ is even for $N=1,2$ because of $-\mathbb{I} \in \Gamma(N)$.
The above modular forms of weight $k$ for $\Gamma(N)$ transform as
\begin{align}
f(\gamma(\tau)) = (c\tau+d)^k \rho(\gamma) f(\tau), \quad
\gamma =
\begin{pmatrix}
a & b \\
c & d
\end{pmatrix}
\in \Gamma, \label{modularformGamma}
\end{align}
under $\Gamma$ transformation, where $\rho$ is a unitary representation of the quotient group $\Gamma'_N \equiv \Gamma/\Gamma(N)$.
Thus, the representation of $\Gamma'_N$, $\rho$, satisfies the following relations,
\begin{align}
\rho(S)^4=[\rho(S)\rho(T)]^3=\rho(T)^N=\mathbb{I}, \quad \rho(S)^2\rho(T) = \rho(T) \rho(S)^2. \label{gamma'Nrep}
\end{align}
Note that since the relation $(-1)^k\rho(-\mathbb{I})=\mathbb{I}$ should be satisfied, it is required that $\rho(-\mathbb{I})=\mathbb{I}$ ($\rho(-\mathbb{I})=-\mathbb{I}$) when $k$ is even (odd).
Consequently, when $k$ is even, $\rho$ becomes a representation of $\Gamma_N \equiv \overline{\Gamma}/\overline{\Gamma}(N)$, where we define $\overline{\Gamma} \equiv \Gamma/\{ \pm \mathbb{I} \}$ and $\overline{\Gamma}(N) \equiv \Gamma(N)/\{ \pm \mathbb{I} \}\ (N=1,2)$ while $\overline{\Gamma} \equiv \Gamma(N)\ (N>2)$.
As mentioned in section \ref{Intro}, $\Gamma_N$ are isomorphic to $\Gamma_2 \simeq S_3$, $\Gamma_3 \simeq A_4$, $\Gamma_4 \simeq S_4$, and $\Gamma_5 \simeq A_5$.
Furthermore, we define the automorphy factor as
\begin{align}
J_k(\gamma, \tau) \equiv (c\tau+d)^k, \quad
\gamma =
\begin{pmatrix}
a & b \\
c & d
\end{pmatrix}
\in \Gamma, \label{automorphyfact}
\end{align}
which satisfies
\begin{align}
J_k(\gamma_1\gamma_2, \tau) = J_k(\gamma_1, \gamma_2(\tau)) J_k(\gamma_2, \tau), \quad \gamma_1, \gamma_2 \in  \Gamma. \label{J}
\end{align}

In the next section, we study the modular symmetry for wavefunctions on the magnetized $T^2$.
In the following analysis, we extend the above discussion on modular forms.


\section{Modular symmetry in the magnetized $T^2$ model}
\label{T2}

We consider ten-dimensional ${\cal N}=1$ supersymmetric Yang-Mills theory, as an effective field theory of magnetized D-brane models of superstring theory, compactified on $T^2 \times T^2 \times T^2$ with non-vanishing magnetic fluxes.
Magnetic fluxes induce degenerate zero-modes corresponding to flavors of quarks and leptons as well as massive modes.
In particular, we focus on one $T^2$ with a magnetic flux, and start with 6D theory.
In this case, the wavefunction of the fermion on 6D space-time, $\lambda(X)$, is decomposed into the wavefunction on 4D space-time, $\psi_n^j(x)$, and the wavefunction on $T^2$, $\psi_n^j(z)$ as follows, 
\begin{align}
\lambda(X) = \sum_{n} \sum_{j} \psi_n^j(x) \otimes \psi_n^j(z), \label{decompose}
\end{align}
where we chose $\psi_n^j(z)$ as the eigenstate of the 2D Dirac operator $i\slash{D}_2$ as
\begin{align}
i\slash{D}_2 \psi_n^j(z) = m_n \psi_n^j(z). \label{eigenT2}
\end{align}
Here, we denote the $n$-th excited and $j$-th degenerate wavefunction as $\psi_n^j$, and also we denote the coordinates of 6D space-time, 4D space-time, and $T^2$ as $X$, $x$, and $z$, respectively.
Then, the 6D action for massless fermion $\lambda(X)$ is reduced to 4D action as
\begin{align}
S &= \int d^6X \overline{\lambda(X)} i \slash{D}_6 \lambda(X) \notag \\
&= \int d^4x \sum_{m,n} \sum_{j,k} \left( \int d^2z \overline{\psi_m^j(z)}  \psi_n^k(z) \right) \overline{\psi_m^j(x)} ( i \slash{D}_4 + m_n ) \psi_n^k(x). \label{action}
\end{align}
In this section, we study the modular symmetry for the wavefunctions on the magnetized $T^2$.


\subsection{Wavefunctions on magnetized $T^2$}
\label{wavT2}

First, we briefly review the wavefunctions on $T^2 \simeq \mathbb{C}/\Lambda$ with $U(1)$ magnetic flux\footnote{The following analysis in this paper can be applied for $U(N)$ magnetic fluxes.}~\cite{Cremades:2004wa}.
The metric on $T^2$ is given by
\begin{align}
ds^2 = 2 h_{\mu\nu} dz^{\mu} d\bar{z}^{\nu}, \quad
h = |e_1|^2
\begin{pmatrix}
0 & \frac{1}{2} \\
\frac{1}{2} & 0
\end{pmatrix}.
\label{metric}
\end{align}
The $U(1)$ magnetic flux on $T^2$,
\begin{align}
F = \frac{i\pi M}{{\rm Im}\tau} dz \wedge d\bar{z},  \label{F}
\end{align}
should satisfy the quantization condition,
\begin{align}
\int_{T^2} F = 2\pi M, \quad M \in \mathbb{Z}. \label{magneticflux}
\end{align}
It is induced from the following vector potential,
\begin{align}
A
&= A_{z} dz + A_{\bar{z}} d\bar{z} \notag \\
&= - \frac{i\pi M}{2{\rm Im}\tau} (\bar{z}+\bar{\zeta}) dz + \frac{i\pi M}{2{\rm Im}\tau} (z+\zeta) d\bar{z} \label{A} \\
&= \frac{\pi M}{{\rm Im}\tau} {\rm Im}\left( (\bar{z}+\bar{\zeta})dz \right), \notag
\end{align}
where $\zeta$ is a Wilson line.
The above vector potential satisfies the following boundary conditions,
\begin{align}
A(z+1) &= A(z) + d\left( \frac{\pi M}{{\rm Im}\tau} {\rm Im}z \right) = A(z) + d\chi_1(z), \label{z1A} \\
A(z+\tau) &= A(z) + d\left( \frac{\pi M}{{\rm Im}\tau} {\rm Im}\bar{\tau}z \right) = A(z) + d\chi_2(z), \label{ztauA}
\end{align}
which correspond to $U(1)$ gauge transformation.

Then, the wavefunctions on the $T^2$ with the above gauge background satisfy the following boundary conditions,
\begin{align}
&\psi(z+1) = e^{i\chi_1(z)} \psi(z) = e^{i\pi M \frac{{\rm Im}(z+\zeta)}{{\rm Im}\tau}} \psi(z), \label{psiz1} \\
&\psi(z+\tau) = e^{i\chi_2(z)} \psi(z) = e^{i\pi M \frac{{\rm Im}\bar{\tau}(z+\zeta)}{{\rm Im}\tau}} \psi(z), \label{psiztau}
\end{align}
where we consider the unit $U(1)$ charge, $q=1$.
Note that the quantization condition $M \in \mathbb{Z}$ originates from the above boundary conditions.
The zero-mode wavefunction of the two-dimensional spinor with charge $q=1$,
\begin{align}
\psi^{M}(z) =
\begin{pmatrix}
\psi^{M}_+(z) \\ \psi^{M}_-(z)
\end{pmatrix}, \quad
\psi^{M}_-(z) = \overline{\psi^{-M}_+(z)}, \label{spinor}
\end{align}
is obtained by solving the zero-mode Dirac equation,
\begin{align}
i (\gamma^z D_z + \gamma^{\bar{z}} D_{\bar{z}}) \psi^{M}(z) &= 0, \label{Dirac}
\end{align}
where $\gamma^{z}, \gamma^{\bar{z}}$ are written by
\begin{align}
\gamma^{z} = \frac{1}{e_1}
\begin{pmatrix}
0 & 2 \\
0 & 0
\end{pmatrix},
\quad
\gamma^{\bar{z}} = \frac{1}{\bar{e}_1}
\begin{pmatrix}
0 & 0 \\
2 & 0
\end{pmatrix},
\label{gammaz}
\end{align}
satisfying $\{ \gamma^{z}, \gamma^{\bar{z}} \} = 2h^{z\bar{z}}$, and we denote the covariant derivative as $D_z = \partial_z - iA_z\ (D_{\bar{z}} = \partial_{\bar{z}} - iA_{\bar{z}})$.
From Eq.~(\ref{gammaz}), each component of Eq.~(\ref{Dirac}) is described by
\begin{align}
i {\cal D} \psi^{M}_+(z) \equiv \frac{2i}{\bar{e}_1} \bar{D}_{\bar{z}} \psi^{M}_+(z) &= 0, \label{psi+EoM} \\
-i \overline{{\cal D}} \psi^{M}_-(z) \equiv \frac{2i}{e_1} D_z \psi^{M}_-(z) &= 0. \label{psi-EoM}
\end{align}
When the magnetic flux $M$ is positive, $\psi^{M}_+(z)$ and $\psi^{-M}_-(z)=\overline{\psi^{M}_+(z)}$ have $|M|$ number of degenerate zero-modes described by
\begin{align}
\psi_0^{j,|M|}(z,\tau)
&= \left(\frac{|M|}{{\cal A}^2}\right)^{1/4} e^{i\pi |M|(z+\zeta) \frac{{\rm Im}(z+\zeta)}{{\rm Im}\tau}} \sum_{l \in \mathbf{Z}} e^{i\pi |M|\tau \left( \frac{j}{|M|}+l \right)^2} e^{2\pi i|M|(z+\zeta) \left( \frac{j}{|M|}+l \right)} \label{psizero} \\
&= \left(\frac{|M|}{{\cal A}^2}\right)^{1/4} e^{i\pi |M|(z+\zeta) \frac{{\rm Im}(z+\zeta)}{{\rm Im}\tau}}
\vartheta
\begin{bmatrix}
\frac{j}{|M|}\\
0
\end{bmatrix}
(|M|z, |M|\tau), 
\notag
\end{align}
with $\forall j \in \mathbb{Z}_{|M|}=\{0,1,2,...,|M|-1\}$,
where ${\cal A} = |e_1|^2 {\rm Im}\tau$ is the area of $T^2$ and $\vartheta$ denotes the Jacobi theta function defined as
\begin{align}
\vartheta
\begin{bmatrix}
a\\
b
\end{bmatrix}
(\nu, \tau)
=
\sum_{l\in \mathbb{Z}}
e^{\pi i (a+l)^2\tau}
e^{2\pi i (a+l)(\nu+b)}.
\end{align}
Similarly, when $M$ is negative,  $\psi^{M}_-(z)$ has $|M|$ degenerate zero-modes, whose wavefuctions are the same as
the above.
Thus, we can realize a chiral theory.

Furthermore, the wavefunctions of the $n$-th excited-modes~\cite{Hamada:2012wj}, whose squared masses are $m_n^2=\frac{4\pi M}{{\cal A}}n$, can be described by
\begin{align}
\psi_n^{j,|M|}(z, \tau)
=& \frac{1}{\sqrt{n!}} \left( a^{\dagger} \right)^n \psi_0^{j,|M|}(z, \tau) \label{psin} \\
=& \frac{1}{\sqrt{n!}} \left( \frac{1}{\sqrt{2}} \right)^n \left( \frac{|M|}{{\cal A}^2} \right)^{1/4} e^{i\pi |M|(z+\zeta)\frac{{\rm Im}(z+\zeta)}{{\rm Im}\tau}} \notag \\
\times& \sum_{l \in \mathbf{Z}} e^{i\pi |M|\tau \left( \frac{j}{|M|}+l \right)^2} e^{2\pi i|M|(z+\zeta) \left( \frac{j}{|M|}+l \right)} H_{n} \left( \sqrt{2\pi |M|{\rm Im}\tau} \left( \frac{{\rm Im}(z+\zeta)}{{\rm Im}\tau}+\frac{j}{|M|}+l \right) \right), \notag
\end{align}
where we use the creation and annihilation operators,
\begin{align}
a^{\dagger} =  \sqrt{\frac{{\cal A}}{4\pi |M|}} \bar{{\cal D}}, \quad a = \sqrt{\frac{{\cal A}}{4\pi |M|}} {\cal D}, \label{creanni}
\end{align}
which satisfy $[a,a^{\dagger}]=1$, and $H_n(x)$ is the Hermite function.
We note that the wavefunctions in Eqs.~(\ref{psizero}) and (\ref{psin}) are normalized by
\begin{align}
\int_{T^2} dzd\bar{z} \overline{\psi_m^{j,|M|}(z)} \psi_n^{k,|M|}(z) = (2{\rm Im}\tau)^{-1/2} \delta_{j,k} \delta_{m,n}. \label{normalize}
\end{align}
From Eqs.~(\ref{normalize}) and (\ref{action}), we can obtain the following 4D kinetic terms,
\begin{align}
S_K = \int d^4x \sum_{n=0}^{\infty} \sum_{j=0}^{|M|-1} \frac{\overline{\psi_n^{j,|M|}(x)} i \slash{D}_4 \psi_n^{j,|M|}(x)}{(-i\tau+i\bar{\tau})^{1/2}}, \label{4Dkinetic}
\end{align}
which means that the wavefunctions on the 4D space-time, $\psi_n^{j,|M|}(x)$, have modular weight $-k=-1/2$.
Thus, the modular symmetry in the 4D low-energy effective field theory is determined by behaviors of wavefunctions 
on the magnetized $T^2$. 
In the next section, we study the modular symmetry on the magnetized $T^2$.
Before ending this section, we also note that the wavefunctions in Eqs.~(\ref{psizero}) and (\ref{psin})  satisfy the following relation,
\begin{align}
\psi_n^{j,|M|}(-z, \tau) = \psi_n^{|M|-j,|M|}(z, \tau). \label{-z}
\end{align}


\subsection{Modular symmetry in the magnetized $T^2$ model}
\label{modularT2}

Here, we study how the fields on the magnetized $T^2$ are transformed under the modular transformation, Eq.~(\ref{ztauSandT}).

The transformation of the Wilson line $\zeta$ is the same as the coordinate $z$, i.e. $\Gamma \ni \gamma: \zeta \rightarrow \zeta/(c\tau+d)$.
The fields $F$ in Eq.~(\ref{F}) and $A$ in Eq.~(\ref{A}) are modular invariant.
The equations of motions for $\psi^{M}(z)$ with any excited-modes, including zero-modes, are also modular invariant.
On the other hand, while the boundary conditions for $\psi^{M}(z)$ in Eqs.~(\ref{psiz1}) and (\ref{psiztau}) are consistent with the $S$ transformation, they are consistent with the $T$ transformation only if $M$ is even.
In general, the boundary conditions are consistent with the modular transformation in Eqs.~(\ref{zSL2Z}) and (\ref{tauSL2Z}) only if $M$ is even or both $ab$ and $cd$ are even, where $a,b,c,d$ are elements of $\gamma \in \Gamma_{1,2} \subset \Gamma$. (See Ref.~\cite{Ohki:2020bpo}.)
Here, we focus on the models with $M$=even.

The wavefunctions of the $n$-th excited-modes in Eq.~(\ref{psin}), including the zero-modes in Eq.~(\ref{psizero}), are transformed as
\begin{align}
&S: \psi_n^{j,|M|}(z, \tau) \rightarrow \psi_n^{j,|M|}\left( -\frac{z}{\tau}, -\frac{1}{\tau} \right) = (-\tau)^{1/2} \sum_{k=0}^{|M|-1} e^{i\pi /4} \frac{1}{\sqrt{|M|}} e^{2\pi i \frac{jk}{|M|}} \psi_n^{k,|M|}(z, \tau), \label{psiS} \\
&T: \psi_n^{j,|M|}(z, \tau) \rightarrow \quad \psi_n^{j,|M|}(z, \tau+1) \ = e^{i\pi \frac{j^2}{|M|}} \psi_n^{j,|M|}(z, \tau), \label{psiT}
\end{align}
under the modular transformation, Eq.~(\ref{ztauSandT}).
Note that the creation operator is modular invariant and commutative with the above coefficients.
Thus, the wavefunctions transform like modular forms of weight $1/2$.
It is consistent with Eq.~(\ref{4Dkinetic}).
Modular forms of  weight $1/2$ are relevant to the double covering group of $\Gamma = SL(2,\mathbb{Z})$, 
$\widetilde{\Gamma} \equiv \widetilde{SL}(2,\mathbb{Z})$.
(See e.g. \cite{Koblitz:1984,shimura,Duncan:2018wbw,Kubota,Budden}.)
The double covering group $\widetilde{\Gamma} \equiv \widetilde{SL}(2,\mathbb{Z})$ is defined by 
\begin{align}
\widetilde{\Gamma} \equiv \left\{ [\gamma, \epsilon] \bigl| \gamma \in \Gamma, \ \epsilon \in \{ \pm 1 \} \right\}. \label{Gammatilde}
\end{align}
The multiplication of arbitrary two elements, $[\gamma_1, \epsilon_1],[\gamma_2, \epsilon_2] \in \widetilde{\Gamma}$, is defined by
\begin{align}
[\gamma_1, \epsilon_1] [\gamma_2, \epsilon_2] = [\gamma_1\gamma_2, A(\gamma_1,\gamma_2)\epsilon_1\epsilon_2], \label{Gammatildemultirule}
\end{align}
where $A(\gamma_1, \gamma_2)$ is called Kubota's twisted 2-cocycle\cite{Kubota} for $\Gamma$, defined as follows.
We first introduce Kubota's function $\chi: \Gamma \rightarrow \mathbb{Z}$, defined by
\begin{align}
\chi(\gamma) = \left\{
\begin{array}{cc}
c, & (c \neq 0) \\
d, & (c = 0)
\end{array} \right. .
\label{chigamma}
\end{align}
We also introduce the Hirbert symbol, defined by
\begin{align}
(a,b)_H = \left\{
\begin{array}{cc}
-1, & (a<0\ {\rm and}\ b<0) \\
1, & ({\rm otherwise})
\end{array} \right. .
\end{align}
Then, $A(\gamma_1, \gamma_2)$ is defined by
\begin{align}
A(\gamma_1,\gamma_2) =
\left( \frac{\chi(\gamma_1\gamma_2)}{\chi(\gamma_1)}, \frac{\chi(\gamma_1\gamma_2)}{\chi(\gamma_2)} \right)_H. \label{Aalphabeta}
\end{align}
Actually, it satisfies the following cocycle relation,
\begin{align}
A(\gamma_1,\gamma_2)A(\gamma_1\gamma_2,\gamma_3) = A(\gamma_1,\gamma_2\gamma_3)A(\gamma_2,\gamma_3). \label{cocycle}
\end{align}
Here, we set
\begin{align}
\mathbb{I} \equiv [\mathbb{I},1], \quad \widetilde{S} \equiv [S,1], \quad \widetilde{T} \equiv [T,1]. \label{IST}
\end{align}
They satisfy the following algebraic relations,
\begin{align}
\widetilde{S}^2=[-\mathbb{I},1] \equiv \widetilde{Z}, \quad \widetilde{S}^4=(\widetilde{S}\widetilde{T})^3=[\mathbb{I}, -1]=\widetilde{Z}^2, \quad \widetilde{S}^8=(\widetilde{S}\widetilde{T})^6=[\mathbb{I},1]=\mathbb{I}=\widetilde{Z}^4. \label{tildealgebrarel}
\end{align}
Note that inverses of $\widetilde{S}, \widetilde{T}, \widetilde{Z}$ are written by 
\begin{align}
\widetilde{S}^{-1} = [S^{-1},1], \quad \widetilde{T}^{-1} = [T^{-1}, 1], \quad \widetilde{Z} = [-\mathbb{I}, -1]. \label{inverseSTZ}
\end{align}
Hereafter, we often denote an element of $\widetilde{\Gamma}$, $[\gamma,\epsilon]$, as $\widetilde{\gamma}$, where $\widetilde{\gamma}$ is, in general, independent of $\gamma$.

Due to the above extension by $\epsilon \in \{ \pm 1\}$, the definition of the automorphy factor in Eq.(\ref{automorphyfact}) is also extended by $\epsilon \in \{ \pm 1\}$ as follows,
\begin{align}
\widetilde{J}_{k/2}(\widetilde{\gamma}, \tau) \equiv \epsilon^{k} J_{k/2}(\gamma,\tau) = \epsilon^{k} (c\tau+d)^{k/2}, \quad k \in \mathbb{Z}, \quad
\widetilde{\gamma} = 
\left[\gamma=
\begin{pmatrix}
a & b \\
c & d
\end{pmatrix}, \epsilon \right] \in \tilde{\Gamma}, \label{Jtilde}
\end{align}
where we take $(-1)^{k/2} = e^{-i\pi k/2}$.
From Eqs.~(\ref{J}) and (\ref{Gammatildemultirule}), Eq.~(\ref{Jtilde}) satisfies the following relation,
\begin{align}
\widetilde{J}_{k/2}(\widetilde{\gamma}_1\widetilde{\gamma}_2, \tau) = \left( A(\gamma_1,\gamma_2) \right)^{k} \widetilde{J}_{k/2}(\widetilde{\gamma}_1, \widetilde{\gamma}_2(\tau)) \widetilde{J}_{k/2}(\widetilde{\gamma}_2, \tau), \quad \widetilde{\gamma}_1=[\gamma_1,\epsilon_1], \widetilde{\gamma}_2=[\gamma_2,\epsilon_2] \in  \widetilde{\Gamma}, \label{Jtilderel}
\end{align}
where the extension by $\epsilon \in \{ \pm 1 \}$ does not affect the modular transformation, i.e.~ $\widetilde{\gamma}(z,\tau)=\gamma(z,\tau)$.
It allows us to study modular forms of weight $k/2$, where $k$ is integer.
Considering the above extension, the wavefunctions on the magnetized $T^2$ transform as
\begin{align}
\psi_n^{j,|M|}(\widetilde{\gamma}(z,\tau)) &= \widetilde{J}_{1/2}(\widetilde{\gamma}, \tau) \sum_{k=0}^{|M|-1} \rho(\widetilde{\gamma})_{jk} \psi_n^{k,|M|}(z,\tau), \quad \widetilde{\gamma} \in \widetilde{\Gamma}, \label{wavemodularform} \\
\rho(\widetilde{S})_{jk} &= e^{i\pi/4} \frac{1}{\sqrt{|M|}} e^{2\pi i\frac{jk}{|M|}}, \quad
\rho(\widetilde{T})_{jk} = e^{i\pi \frac{j^2}{|M|}} \delta_{j,k}, \label{rhoSandT}
\end{align}
under the modular transformation.
Note that $\widetilde{Z}(z,\tau)=(-z,\tau)$ and $\widetilde{Z}^2(z,\tau)=(z,\tau)$ require $\widetilde{J}_{1/2}(\widetilde{Z},\tau)\rho(\widetilde{Z})=\delta_{|M|-j,k}$ and $\widetilde{J}_{1/2}(\widetilde{Z}^2,\tau)\rho(\widetilde{Z})^2=\delta_{j,k}$, respectively.
Actually, we can check that the following relations,
\begin{align}
\widetilde{J}_{1/2}(\widetilde{Z},\tau) = \widetilde{J}_{1/2}(\widetilde{S}^2,\tau) = -i,
\quad&
\rho(\widetilde{Z})_{jk} = \rho(\widetilde{S})^2_{jk} = i \delta_{|M|-j,k}, \label{Z} \\
\widetilde{J}_{1/2}(\widetilde{Z}^2,\tau) = \widetilde{J}_{1/2}(\widetilde{S}^4,\tau) = \widetilde{J}_{1/2}((\widetilde{S}\widetilde{T})^3,\tau) = -1,
\quad&
\rho(\widetilde{Z})^2_{jk} = \rho(\widetilde{S})^4_{jk} = [\rho(\widetilde{S})\rho(\widetilde{T})]^3_{jk} = -\delta_{j,k}, \label{Z2} \\
\widetilde{J}_{1/2}(\widetilde{Z}^4,\tau) = \widetilde{J}_{1/2}(\widetilde{S}^8,\tau) = \widetilde{J}_{1/2}((\widetilde{S}\widetilde{T})^6,\tau) = 1,
\quad&
\rho(\widetilde{Z})^4_{jk} = \rho(\widetilde{S})^8_{jk} = [\rho(\widetilde{S})\rho(\widetilde{T})]^6_{jk} = \delta_{j,k}, \label{Z4} \\
\widetilde{J}_{1/2}(\widetilde{T}^n,\tau) = 1, \ ^{\forall} n \in \mathbb{Z}, \quad& \rho(\widetilde{T})^{2|M|}_{jk} = \delta_{j,k}, \label{Tn} \\
& \rho(\widetilde{Z})^n\rho(\widetilde{T}) = \rho(\widetilde{T})\rho(\widetilde{Z})^n, \ n=1,2,3, \label{ZT}
\end{align}
are satisfied\footnote{The following relations are also satisfied
\begin{align}
[\rho(\widetilde{S})\rho(\widetilde{T})^{-1}]^3_{jk}=i\delta_{|M|-j,k}, \quad [\rho(\widetilde{S})\rho(\widetilde{T})^{-1}]^6_{jk}=-\delta_{j,k}, \quad [\rho(\widetilde{S})\rho(\widetilde{T})^{-1}]^{12}_{jk}=\delta_{j,k}. \label{ST-1}
\end{align}}.
Therefore, the wavefunctions on the magnetized $T^2$ transform under the modular transformation like modular forms of weight $1/2$ for $\widetilde{\Gamma}(2|M|)$~\footnote{According to Ref.~\cite{shimura}, $\widetilde{SL}(2,\mathbb{Z}) \rightarrow SL(2,\mathbb{Z})$ can be split on $\Gamma(2|M|)$ since $2|M| \in 4\mathbb{Z}$.}, which is the normal subgroup of $\widetilde{\Gamma}$, defined as
\begin{align}
\widetilde{\Gamma}(2|M|) \equiv \{ [h, \epsilon] \in \widetilde{\Gamma} | h \in \Gamma(2|M|), \epsilon=1 \}. \label{gammatilde2M}
\end{align}
Then $\rho$ is a unitary representation of the quotient group $\widetilde{\Gamma}'_{2|M|} \equiv \widetilde{\Gamma}/\widetilde{\Gamma}(2|M|)$.
That is, the group generated by $\rho$ is homomorphic with $\widetilde{\Gamma}'_{2|M|}$.

For example, when $M=2$, the $S$ and $T$ transformations are represented as
\begin{align}
	\rho(\widetilde{S})={e^{i\pi /4} \over \sqrt{2}}\begin{pmatrix} 1 & 1 \\ 1 & -1\end{pmatrix},\quad
	\rho(\widetilde{T})=\begin{pmatrix} 1 & 0 \\ 0 & i\end{pmatrix}.  \label{M2rep}
\end{align}
They generate the group $G^2$ whose order is $96$, and it is isomorphic to 
\begin{equation}
	G^2 \simeq T' \rtimes Z_4. \label{M2group}
\end{equation}
When $M=4$, the $S$ and $T$ transformations are represented as
\begin{align}
	\rho(\widetilde{S})={e^{i\pi /4} \over 2}\begin{pmatrix} 1 & 1 & 1 & 1 \\ 1 & i & -1 & -i \\ 1 & -1 & 1 & -1 \\ 1 & -i & -1 & i \\ \end{pmatrix},\quad
	\rho(\widetilde{T})=\begin{pmatrix} 1 & 0 & 0 & 0 \\ 0 & e^{i\pi /4} & 0 & 0 \\ 0 & 0 & -1 & 0 \\ 0 & 0 & 0 &  e^{i\pi /4} \\ \end{pmatrix}. \label{M4rep}
\end{align}
They generate the group $G^4$, whose order is $384$, and it is isomorphic to 
\begin{align}
	G^4 &\simeq \Delta(48) \rtimes Z_8. \label{M4group}
\end{align}
In Appendix~\ref{gCP}, we study the extension for the generalized $CP$ symmetry with the modular symmetry on the magnetized $T^2$.

So far, we have considered the Wilson line $\zeta$, which transforms as $\zeta \rightarrow \zeta/(c\tau +d)$. 
In that case, the modular transformation is restrictive due to the consistency with the boundary conditions for $\psi^{M}(z)$.
Before ending this section, we comment about another possibility.
In particular, if a Wilson line $\zeta$ is also changed to $\zeta+1$, which is the gauge transformation, simultaneously with the $T$ transformation, the boundary conditions for $\psi^{M}(z)$ are consistent with the modular transformation even if $M$ is odd, although the equations of motions for $\psi^{M}(z)$ are modified.
In this case, the zero-mode wavefunction for $j$ after the $T$ transformation can be expanded by the all excited-mode wavefunctions for $j$ before the $T$ transformation as follows,
\begin{align}
T: \psi_0^{j,|M|}(z+\zeta, \tau) \rightarrow& \psi_0^{j,|M|}(z+\zeta+1, \tau+1) \notag \\
&= (-1)^j e^{i\pi \frac{j^2}{|M|}} e^{-\frac{\pi |M|}{8{\rm Im}\tau}} \sum_{n=0}^{\infty} \frac{1}{\sqrt{n!}} \left(i\sqrt{\frac{\pi |M|}{4{\rm Im}\tau}}\right)^n \psi_{n}^{j,|M|}(z+\zeta,\tau), \label{Twilsonzeromassive}
\end{align}
where we use the following the generating function of the Hermite function,
\begin{align}
e^{-y^2+2xy} = \sum_{n=0} H_n(x) \frac{y^n}{n!}. \label{Hermitobokansu}
\end{align}
The detail calculation is shown in Appendix~\ref{withgauge}.
Similarly, the $n$-th excited-mode wavefunction for $j$ after the $T$ transformation can be also expanded by the all excited-mode wavefunctions for $j$ before the $T$ transformation.

In this section, we have discussed the modular symmetry on magnetized $T^2$.
In the following sections, we study the modular symmetry on various magnetized $T^2/\mathbb{Z}_N$ orbifolds.


\section{Modular symmetry in magnetized $T^2/\mathbb{Z}_N$ twisted orbifold models}
\label{T2ZNtwist}

In this section, we study the modular symmetry of the wavefunctions on the magnetized $T^2/\mathbb{Z}_N$ twisted orbifolds~\cite{Abe:2008fi,Abe:2013bca,Abe:2014noa,Kobayashi:2017dyu}.
Here and hereafter, we often omit the KK index $n$, because each KK level satisfies the same relations in what follows.
For simplicity, we do not introduce non-vanishing discrete Wilson lines, although we can discuss models with non-vanishing 
Wilson lines similarly.
The $T^2/\mathbb{Z}_N$ twisted orbifold can be obtained by further identifying the points on $T^2 \simeq \mathbb{C}/\Lambda$ which are rotated by $\alpha^k_N \equiv e^{2\pi ik/N}; ^{\forall} k \in \mathbb{Z}_N=\{0,1,2,...,N-1\}$.
That is the $ \mathbb{Z}_N$ twist,  i.e. $(\alpha^k_N)^N=1$.
Hence, a lattice point, except for the origin, should move to another lattice point after any $\mathbb{Z}_N$ twist. 
It allows only if $N=2,3,4,6$.
Moreover, the modulus $\tau=e_2/e_1$ should be fixed to be $\tau=\alpha_N=e^{2\pi i/N}$ for $N=3,4,6$, although any $\tau$ is allowed for $N=2$.
It means that only $ST, S, ST^{-1}$ transformations of the modular transformations are consistent for $N=3,4,6$, respectively, while there remains the full modular symmetry for $N=2$.

The wavefunction on the magnetized $T^2/\mathbb{Z}_N$ twisted orbifold, $\psi_{T^2/\mathbb{Z}^m_N}^{j,|M|}(z)$ must satisfy 
the following boundary condition,
\begin{align}
\psi_{T^2/\mathbb{Z}^m_N}^{j,|M|}(\alpha_N z) = \alpha^m_N \psi_{T^2/\mathbb{Z}^m_N}^{j,|M|}(z), \quad m \in \mathbb{Z}_N. \label{twistT2ZNbase}
\end{align}
Hence, such wavefunctions 
can be written by linear combinations of wavefunctions on the magnetized $T^2$ as
\begin{align}
\psi_{T^2/\mathbb{Z}^m_N}^{j,|M|}(z) &= {\cal N}_N^t \sum_{k=0}^{N-1} (\alpha^m_N)^{-k} \psi_{T^2}^{j,|M|}(\alpha^k_N z),
\label{T2ZNexpT2}
\end{align}
where ${\cal N}_N^t$ is the normalization factor determined by remaining Eq.~(\ref{normalize}).
Furthermore, $\psi_{T^2}^{j,|M|}(\alpha^k_N z)$ satisfies the same equation of motion as $\psi_{T^2}^{j,|M|}(z)$.
In addition, if $\psi_{T^2}^{j,|M|}(\alpha^k_N z)$ also satisfies the same boundary condition as $\psi_{T^2}^{j,|M|}(z)$,
$\psi_{T^2}^{j,|M|}(\alpha^k_N z)$ can be expanded by the same excited-mode of $\psi_{T^2}^{j,|M|}(z)$.

First, we consider the magnetized $T^2/\mathbb{Z}_2$ twisted orbifold.
In this case, since the wavefunction $\psi_n^{j,|M|}(\alpha_2 z, \tau)$ satisfies the same boundary conditions, Eqs.~(\ref{psiz1}) and (\ref{psiztau}), $\psi_n^{j,|M|}(\alpha_2 z, \tau)$ can be expressed by $\psi_n^{j,|M|}(z, \tau)$, i.e.~Eq.~(\ref{-z}).
Therefore, the wavefunction on the magnetized $T^2/\mathbb{Z}_2$ twisted orbifold basis, $\psi_{T^2/\mathbb{Z}^m_2}^{j,|M|}(z)$, can be written by linear combinations of wavefunctions on the magnetized $T^2$ basis, $\psi_{T^2}^{j,|M|}(z)$, as
\begin{align}
\psi^{j,|M|}_{T^2/\mathbb{Z}_2^{m}}(z,\tau) &=  {\cal N}_2^t \sum_{k=0}^{|M|-1} \left( \delta_{j,k} + (-1)^m \delta_{|M|-j,k} \right) \psi^{k,|M|}_{T^2}(z,\tau), \label{T2Z2expT2}
\end{align}
where the normalization factor ${\cal N}_2^t$ is determined by ${\cal N}_2^t=1,1/2,$ and $1/\sqrt{2}$ for $j=0,|M|/2,$ and the others, respectively.
Note that there are no $\mathbb{Z}_2$-odd modes, $\psi^{j,|M|}_{T^2/\mathbb{Z}_2^{1}}(z,\tau)$, for $j=0, |M|/2$.
When $M$ is even, the numbers of $\mathbb{Z}_2$-even $(m=0)$ and -odd $(m=1)$ modes are $(|M|/2+1)$ and $(|M|/2-1)$, respectively\footnote{
When $M$ is odd, the numbers of $\mathbb{Z}_2$-even and $\mathbb{Z}_2$-odd modes are $((|M|-1)/2+1)$ and $((|M|-1)/2)$, respectively.}.
On the $T^2/\mathbb{Z}_2$ twisted orbifold basis, Eq.~(\ref{rhoSandT}) is deformed by
\begin{align}
\rho_{T^2/\mathbb{Z}_2^{0}}(\widetilde{S})_{jk} &= e^{i\pi/4} \frac{2}{\sqrt{|M|}} \cos (2\pi jk/|M|), \quad
\rho_{T^2/\mathbb{Z}_2^{0}}(\widetilde{T})_{jk} = e^{i\pi \frac{j^2}{|M|}} \delta_{j,k}, \label{rho+SandT} \\
\rho_{T^2/\mathbb{Z}_2^{1}}(\widetilde{S})_{jk} &= e^{i\pi/4} \frac{2i}{\sqrt{|M|}} \sin (2\pi jk/|M|), \quad
\rho_{T^2/\mathbb{Z}_2^{1}}(\widetilde{T})_{jk} = e^{i\pi \frac{j^2}{|M|}} \delta_{j,k}, \label{rho+SandT}
\end{align}
where we need to multiply $\rho_{T^2/\mathbb{Z}_2^{0}}(\widetilde{S})$ further by $1/\sqrt{2}$ when $j$ or $k$ is $0$ or $|M|/2$.
The above deformations induce  deformation of the relation in Eq.~(\ref{Z}) as
\begin{align}
\rho_{T^2/\mathbb{Z}_2^{m}}(\widetilde{Z})_{jk} = \rho_{T^2/\mathbb{Z}_2^{m}}(\widetilde{S})^2_{jk} = (-1)^m i \delta_{j,k}, \label{Zm}
\end{align}
while the other relations are the same as the $T^2$ basis.
Thus, the representations on the $T^2/\mathbb{Z}_2$ twisted orbifold basis satisfy the same algebraic relations as that on the $T^2$ basis, although the dimensions of the representations
are different.
For example, when $M=4$, the wavefunctions on the $T^2/\mathbb{Z}_2$ twisted orbifold basis are expressed as
\begin{align}
\begin{pmatrix}
\psi^{0,4}_{T^2/\mathbb{Z}_2^{0}}(z,\tau) \\ \psi^{1,4}_{T^2/\mathbb{Z}_2^{0}}(z,\tau) \\ \psi^{2,4}_{T^2/\mathbb{Z}_2^{0}}(z,\tau)
\end{pmatrix}
&=
\begin{pmatrix}
\psi^{0,4}_{T^2}(z,\tau) \\ \frac{1}{\sqrt{2}} \left( \psi^{1,4}_{T^2}(z,\tau) + \psi^{3,4}_{T^2}(z,\tau) \right) \\ \psi^{2,4}_{T^2}(z,\tau)
\end{pmatrix}, \label{psiT2Z2M4m0} \\
\psi^{1,4}_{T^2/\mathbb{Z}_2^{1}}(z,\tau) &= \frac{1}{\sqrt{2}} \left( \psi^{1,4}_{T^2}(z,\tau) - \psi^{3,4}_{T^2}(z,\tau) \right). \label{psiT2Z2M4m1}
\end{align}
The representations of the $S$ and $T$ transformations for the $\mathbb{Z}_2$-even modes are expressed as
\begin{align}
	\rho_{T^2/\mathbb{Z}_2^{0}}(\widetilde{S})={e^{i\pi /4} \over 2}\begin{pmatrix} 1 & \sqrt{2} & 1 \\ \sqrt{2} & 0 & -\sqrt{2} \\ 1 & -\sqrt{2} & 1 \\ \end{pmatrix}, \quad
	&\rho_{T^2/\mathbb{Z}_2^{0}}(\widetilde{T})=\begin{pmatrix} 1 & 0 & 0 \\ 0 & e^{i\pi /4} & 0 \\ 0 & 0 & -1 \\ \end{pmatrix}, \label{M4evenrep}
\end{align}
which are the generators of the group $G^4_0$.
The group $G^4_0$ has the order $384$ and is isomorphic to 
\begin{align}
	G^4_0 \simeq \Delta(48) \rtimes Z_8, \label{M4group0}
\end{align}
which is the same as the group on $T^2$ in Eq.(\ref{M4group}).
The above wavefunctions in Eq.~(\ref{psiT2Z2M4m0}) correspond to a triplet under  $G^4_0 \simeq \Delta(48) \rtimes Z_8$.
The representations of the $S$ and $T$ transformations for the $\mathbb{Z}_2$-odd mode, on the other hand, are expressed as
\begin{align}
	\rho_{T^2/\mathbb{Z}_2^{1}}(\widetilde{S})=e^{3\pi i/4}, \quad \rho_{T^2/\mathbb{Z}_2^{1}}(\widetilde{T})=e^{i\pi /4}, \label{M4oddrep}
\end{align}
which are the generators of the group $G^4_1$.
The group $G^4_1$ is nothing but
\begin{align}
	G^4_1 \simeq Z_8, \label{M4group1}
\end{align}
which is a subgroup of $G_0^4 \simeq \Delta(48) \rtimes Z_8$.
The above representation in Eq.~(\ref{psiT2Z2M4m1}) is a representation of this $Z_8$ symmetry and 
it also corresponds to a singlet under  $G_0^4 \simeq \Delta(48) \rtimes Z_8$.

Thus,  the $T^2/\mathbb{Z}_2$ twisted orbifold is consistent with the modular symmetry.
Furthermore, the wavefuncitons on $T^2$ are decomposed into smaller representations 
by $\mathbb{Z}_2$ eigenvalues, even and odd, that is, the  $T^2/\mathbb{Z}_2$ twisted orbifold basis.
For smaller $|M|$, this basis of wavefunctions provide us with irreducible representations of 
$\widetilde{\Gamma}'_{2|M|} \equiv \widetilde{\Gamma}/\widetilde{\Gamma}(2|M|)$.
For larger $|M|$,  wavefunctions on the  $T^2/\mathbb{Z}_2$ twisted orbifold basis could be 
decomposed further.
We will study it in section~\ref{T2Z2twistshift}.

Next, we comment about the other magnetized $T^2/\mathbb{Z}_N$ twisted orbifolds.
In the case of $T^2/\mathbb{Z}_4$, since the wavefunction $\psi_n^{j,|M|}(\alpha_4^k z, \alpha_4)$ satisfies the same boundary condition as $\psi_{n}^{j,|M|}(z,\alpha_4)$, $\psi_n^{j,|M|}(\alpha_4^k z, \alpha_4)$ can be expanded by $\psi_{n}^{j,|M|}(z,\alpha_4)$.
Actually, it can be done by considering $S$ transformation for $\psi_{n}^{j,|M|}(z,\alpha_4)$~\cite{Kobayashi:2017dyu}.
Therefore, the wavefunction on the magnetized $T^2/\mathbb{Z}_4$ twisted orbifold basis, $\psi_{T^2/\mathbb{Z}^m_4}^{j,|M|}(z)$, can be expanded by linear combinations of wavefunctions on the magnetized $T^2$ basis, $\psi_{T^2}^{j,|M|}(z)$, as
\begin{align}
&\psi_{T^2/\mathbb{Z}^m_4}^{j,|M|}(z,\alpha_4) \notag \\
&= {\cal N}_4^t \sum_{k=0}^{|M|-1} \left( (\delta_{j,k} + (-1)^{m} \delta_{|M|-j,k} ) + \frac{e^{-i\pi m/2}}{\sqrt{|M|}} ( e^{2\pi i \frac{jk}{|M|}} + (-1)^m e^{-2\pi i \frac{jk}{|M|}} ) \right) \psi^{k,|M|}_{T^2}(z,\alpha_4). \label{T2Z4expT2}
\end{align}
On the $T^2/\mathbb{Z}_4$ twisted orbifold basis, the representation of $S$ transformation is diagonalized as
\begin{align}
\rho_{T^2/\mathbb{Z}^m_4}(\widetilde{S})_{jk} = e^{i\pi /4} (e^{i\pi /2})^m \delta_{j,k}. \label{rhoSZ4}
\end{align}
That is the $Z_8$ symmetry.

In the case of $T^2/\mathbb{Z}_N$ for $N=3,6$, however, the wavefunction $\psi_n^{j,|M|}(\alpha_N^k z, \alpha_N)$ satisfies the same boundary condition as $\psi_{n}^{j,M}(z,\alpha_N)$ only if $M$ is even.
Thus, when $M$ is even, $\psi_n^{j,|M|}(\alpha_N^k z, \alpha_N)$ can be expanded by $\psi_{n}^{j,|M|}(z,\alpha_N)$.
Actually, it can be done by considering $ST, ST^{-1}$ transformations for $\psi_{n}^{j,|M|}(z,\alpha_N); N=3,6$, respectively~\cite{Kobayashi:2017dyu}.
Therefore, the wavefunction on the magnetized $T^2/\mathbb{Z}_N; N=3,6$ twisted orbifold base, $\psi_{T^2/\mathbb{Z}^m_N}^{j,|M|}(z)$, can be expanded by linear combinations of wavefunctions on the magnetized $T^2$ basis, $\psi_{T^2}^{j,|M|}(z)$, as
\begin{align}
&\psi_{T^2/\mathbb{Z}^m_3}^{j,|M|}(z,\alpha_3) \notag \\
&= {\cal N}_3^t \sum_{k=0}^{|M|-1} \left( \delta_{j,k} - \frac{e^{2\pi im/3}}{\sqrt{|M|}} ( e^{-i\pi /12} e^{2\pi i \frac{jk}{|M|}} e^{i\pi \frac{k^2}{|M|}} + e^{-2\pi im/3} e^{i\pi /12} e^{-i\pi \frac{j^2}{|M|}} e^{-2\pi i \frac{jk}{|M|}} ) \right) \psi^{k,|M|}_{T^2}(z,\alpha_3), \label{T2Z3expT2} \\
&\psi_{T^2/\mathbb{Z}^m_6}^{j,|M|}(z,\alpha_6) \notag \\
&= {\cal N}_6^t \sum_{k=0}^{|M|-1} \Biggl( (\delta_{j,k} + (-1)^{m} \delta_{|M|-j,k} ) \notag \\
&\hspace{0.6cm} + \frac{e^{-i\pi m/3}}{\sqrt{|M|}} \biggl( ( e^{i\pi /12} e^{2\pi i \frac{jk}{|M|}} e^{-i\pi \frac{k^2}{|M|}} + e^{-i\pi m/3} e^{-i\pi /12} e^{i\pi \frac{j^2}{|M|}} e^{2\pi i \frac{jk}{|M|}} ) \notag \\
&\hspace{1.2cm} - (-1)^m ( e^{i\pi /12} e^{-2\pi i \frac{jk}{|M|}} e^{-i\pi \frac{k^2}{|M|}} + e^{-i\pi m/3} e^{-i\pi /12} e^{i\pi \frac{j^2}{|M|}} e^{-2\pi i \frac{jk}{|M|}} ) \biggl) \Biggl) \psi^{k,|M|}_{T^2}(z,\alpha_6). \label{T2Z6expT2}
\end{align}
On the $T^2/\mathbb{Z}_3$ twisted orbifold base, the representation of $ST$ transformation is diagonalized as
\begin{align}
\rho_{T^2/\mathbb{Z}^m_3}(\widetilde{S}\widetilde{T})_{jk} = e^{i\pi /3} (e^{2\pi i/3})^m \delta_{j,k}. \label{rhoSTZ3}
\end{align}
That is the $Z_6$ symmetry.
On the $T^2/\mathbb{Z}_6$ twisted orbifold base, the representation of $ST^{-1}$ transformation is diagonalized as
\begin{align}
\rho_{T^2/\mathbb{Z}^m_3}(\widetilde{S}\widetilde{T}^{-1})_{jk} = e^{i\pi /6} (e^{i\pi /3})^m \delta_{j,k}. \label{rhoSTZ3}
\end{align}
That is the $Z_{12}$ symmetry.
Thus, there remain $Z_{2N}$ symmetries in $\rho(\gamma)$ 
on the magnetized $\mathbb{Z}_N$ twisted orbifolds for $N=3,4,6$.
Remaining $\rho(\gamma)$ represent a spinor representation under 
$\mathbb{Z}_N$ twist.
Obviously, $\rho(\gamma)$ on the $T^2$ and $\mathbb{Z}_2$ bases also 
correspond to spinor representations under the 2D (discrete) rotation.


\section{Modular symmetry in magnetized $T^2/\mathbb{Z}_N$ shifted orbifold models}
\label{T2ZNshift}

In this section, we study the modular symmetry for the wavefunctions on the magnetized $T^2/\mathbb{Z}_N$ shifted orbifolds~\cite{Fujimoto:2013xha}.
The $T^2/\mathbb{Z}_N$ shifted orbifold can be obtained by further identifying the points on $T^2 \simeq \mathbb{C}/\Lambda$ which are shifted by $ke_N^{(m,n)} \equiv k(m+n\tau)/N; ^{\forall} k, ^{\exists} m, ^{\exists} n \in \mathbb{Z}_N=\{0,1,2,...,N-1\}$.
Then, the wavefunctions on the $T^2/\mathbb{Z}_N$ shifted orbifold have to also satisfy  the following boundary condition,
\begin{align}
\psi_{T^2/\mathbb{Z}^\ell_N}^{j,|M|}(z+e_N^{(m,n)}) = \alpha^\ell_N e^{i\chi_N^{(m,n)}(z)} \psi_{T^2/\mathbb{Z}^\ell_N}^{j,|M|}(z) = e^{2\pi i\ell/N} e^{i\pi |M| \left( \frac{{\rm Im}\bar{e}_N^{(m,n)}(z+\zeta)}{{\rm Im}\tau}+\frac{mn}{N} \right)} \psi_{T^2/\mathbb{Z}^\ell_N}^{j,|M|}(z), \label{psizmn}
\end{align}
with $\ell \in \mathbb{Z}_N$, 
which is consistent with the boundary condition for $z \rightarrow z+m+n\tau$, in addition to Eqs.~(\ref{psiz1}) and (\ref{psiztau}).
Furthermore, these boundary conditions constrain the magnetic flux $M$ to be $M=Nt, t \in \mathbb{Z}$.
The above wavefunction can be written by linear combinations of wavefunctions on the magnetized $T^2$  as
\begin{align}
\psi_{T^2/\mathbb{Z}^\ell_N}^{j,|M|}(z) &= {\cal N}_N^s \sum_{k=0}^{N-1} (\alpha^\ell_N)^{-k} e^{-ik\chi_N^{(m,n)}(z)} \psi_{T^2}^{j,|M|}(z+ke_N^{(m,n)}),
\label{shiftT2ZNexpT2}
\end{align}
where ${\cal N}_N^s$ is the normalization factor determined by remaining Eq.~(\ref{normalize}).
Furthermore, since $e^{-ik\chi_N^{(m,n)}(z,\tau)} \psi_{T^2}^{j,|M|}(z+ke_N^{(m,n)})$ satisfies the same equation of motion as $\psi_{T^2}^{j,|M|}(z)$ and also satisfies the same boundary condition as $\psi_{T^2}^{j,|M|}(z)$, 
one can expand 
$e^{-ik\chi_N^{(m,n)}(z,\tau)} \psi_{T^2}^{j,|M|}(z+ke_N^{(m,n)})$ by the same excited-mode of $\psi_{T^2}^{j,|M|}(z)$.
Then, the wavefunction on the magnetized $T^2/\mathbb{Z}_N$ shifted orbifold, $\psi_{T^2/\mathbb{Z}^\ell_N}^{j,|M|}(z)$, can be expanded by linear combinations of wavefunctions on the magnetized $T^2$, $\psi_{T^2}^{j,|M|}(z)$, as
\begin{align}
\psi_{T^2/\mathbb{Z}^\ell_N}^{j,|M|}(z,\tau) = {\cal N}_N^s \sum_{k=0}^{N-1} e^{-2\pi ik(\ell-mj)/N} e^{-i\pi k(N-k)mn|t|/N} \psi_{T^2}^{j+kn|t|,|M|}(z,\tau), \label{shiftT2ZNexp}
\end{align}
which can be obtained from Eqs.~(\ref{psizero}) and (\ref{psin})  directly.
The normalization factor ${\cal N}_N^s$ is determined as $ {\cal N}_N^s=1/\sqrt{N}$ for $n \neq 0$ or ${\cal N}_N^s=1/N$ for $n=0$.

We discuss the modular symmetry on the magnetized $T^2/\mathbb{Z}_N$ shifted orbifolds.
There is the modular symmetry on the $T^2/\mathbb{Z}_N$ shifted orbifold only if the points on $T^2 \simeq \mathbb{C}/\Lambda$ which are shifted by $e_N^{(m,n)}=(m+n\tau)/N;  ^{\forall} m, ^{\forall} n \in \mathbb{Z}_N$ are further identified.
Hereafter, we call this $T^2/\mathbb{Z}_N$ shifted orbifold the full $T^2/\mathbb{Z}_N$ shifted orbifold.
The full $T^2/\mathbb{Z}_N$ shifted orbifold with magnetic flux $M$ corresponds to $T^{2'} \simeq \mathbb{C}/\Lambda', \Lambda' \equiv \Lambda/N$ with magnetic flux $M/N^2$.
The boundary conditions for the wavefunctions on the magnetized full $T^2/\mathbb{Z}_N$ shifted orbifold are written by,
\begin{align}
\psi_{T^2/\mathbb{Z}^{(\ell_1,\ell_2)}_N}^{j,|M|}(z+e_N^{(1,0)}) &= \alpha^{\ell_1}_N e^{i\chi_N^{(1,0)}(z)} \psi_{T^2/\mathbb{Z}^{(\ell_1,\ell_2)}_N}^{j,|M|}(z), \label{psiz10} \\
\psi_{T^2/\mathbb{Z}^{(\ell_1,\ell_2)}_N}^{j,|M|}(z+e_N^{(0,1)}) &= \alpha^{\ell_2}_N e^{i\chi_N^{(0,1)}(z)} \psi_{T^2/\mathbb{Z}^{(\ell_1,\ell_2)}_N}^{j,|M|}(z). \label{psiz01}
\end{align}
The above boundary conditions are consistent with Eqs.~(\ref{psiz1}), (\ref{psiztau}), and (\ref{psizmn}) for $^{\forall} m, ^{\forall} n \in \mathbb{Z}_N$, where we denote $\ell$ in Eq.~(\ref{psizmn}) as $\ell^{(m,n)}$, determined by $\ell^{(m,n)}\equiv m\ell_1+n\ell_2\ ({\rm mod}\ N )$.
From the above boundary conditions, we obtain $s \equiv M/N^2 \in \mathbb{Z}$.
The eigenfunctions for $^{\forall} e^{(m,n)}_N$-shifts which satisfy the above boundary conditions are expressed as
\begin{align}
\Psi_{T^2/\mathbb{Z}^{(\ell_1,\ell_2)}_N}^{r,|s|}(z,\tau) \equiv \psi_{T^2/\mathbb{Z}^{(\ell_1,\ell_2)}_N}^{j,|M|}(z,\tau) = \frac{1}{\sqrt{N}} \sum_{k=0}^{N-1} e^{-2\pi ik\ell_2/N} \psi_{T^2}^{j+kN|s|,|M|}(z,\tau), \label{shiftT2ZNexpall} \\
M=N^2s, \ s \in \mathbb{Z}, \ j=Nr+\ell_1 \in \mathbb{Z}_{N|s|}, \ r \in \mathbb{Z}_{|s|}, \ \ell_1, \ell_2 \in \mathbb{Z}_N, \notag
\end{align}
where we note that $\ell_1 \equiv j\ ({\rm mod}\ N )$.
Furthermore, when we consider $s$=even,
the boundary conditions, Eqs.~(\ref{psiz10}) and (\ref{psiz01}), are consistent with the modular transformation.
On this full $T^2/\mathbb{Z}_N$ shifted orbifold basis, Eq.~(\ref{rhoSandT}) is deformed as
\begin{align}
&\Psi_{T^2/\mathbb{Z}^{(\ell_1,\ell_2)}_N}^{r,|s|}(\widetilde{\gamma}(z,\tau)) = \widetilde{J}_{1/2}(\widetilde{\gamma}, \tau) \sum_{r'=0}^{|s|-1} \sum_{\ell'_1,\ell'_2=0}^{N-1} \rho_{T^2/\mathbb{Z}_N^{(\ell_1,\ell_2)}}(\widetilde{\gamma})_{rr',(\ell_1,\ell_2)(\ell'_1,\ell'_2)} \Psi_{T^2/\mathbb{Z}^{(\ell'_1,\ell'_2)}_N}^{r',|s|}(z,\tau),
\end{align}
for $ \widetilde{\gamma} \in \widetilde{\Gamma}$, where 
\begin{align} 
&\rho_{T^2/\mathbb{Z}_N^{(\ell_1,\ell_2)}}(\widetilde{S})_{rr',(\ell_1,\ell_2)(\ell'_1,\ell'_2)} = e^{i\pi/4} \frac{1}{\sqrt{|s|}} e^{2\pi i \left(\frac{\ell_1}{N}+r\right) \left(\frac{\ell'_1}{N}+r'\right)/|s|} \delta_{\ell_2,\ell'_1} \delta_{N-\ell_1,\ell'_2}, \label{rhoshiftZNS} \\
&\rho_{T^2/\mathbb{Z}_N^{(\ell_1,\ell_2)}}(\widetilde{T})_{rr',(\ell_1,\ell_2)(\ell'_1,\ell'_2)} = e^{i\pi \left(\frac{\ell_1}{N}+r\right)^2/|s|} \delta_{r,r'} \delta_{\ell_1,\ell'_1} \delta_{\ell_2-\ell_1,\ell'_2}. \label{rhoshiftZNT}
\end{align}
The above deformations induce the deformation of the relation in Eq.~(\ref{Z}) as
\begin{align}
&\rho_{T^2/\mathbb{Z}_N^{(\ell_1,\ell_2)}}(\widetilde{Z})_{rr',(\ell_1,\ell_2)(\ell'_1,\ell'_2)} &= \rho_{T^2/\mathbb{Z}_N^{(\ell_1,\ell_2)}}(\widetilde{S})_{rr',(\ell_1,\ell_2)(\ell'_1,\ell'_2)}^2 = e^{2\pi i\ell_2/N} i \delta_{|s|-r-1,r'} \delta_{N-\ell_1,\ell'_1} \delta_{N-\ell_2,\ell'_2}. \label{ZshiftT2ZN}
\end{align}
We should modify several terms in the following particular case.
Since $N-\ell_1 \equiv \ell_1\ ({\rm mod}\ N)$ is satisfied when $\ell_1=0$ or $N=2$, $\delta_{|s|-r-1,r'}$ should be modified into $\delta_{|s|-r,r'}$ when $\ell_1=0$ or $\delta_{|s|-r-\ell_1,r'}$ when $N=2$.
Furthermore, when $r=0$ in addition to $\ell_1=0$ or $N=2$, $e^{2\pi i\ell_2/N}$ does not appear even if $\ell_2 \neq 0$.
Note that Eq.~(\ref{ZshiftT2ZN}) leads to the following relation\footnote{The following calculation is useful to confirm Eq.~(\ref{-zshiftT2ZN}),
\begin{align}
e^{-2\pi ik\ell_2/N} \psi^{(Nr+\ell_1)+kN|s|,N^2|s|}_{T^2}(-z)
&= e^{-2\pi ik\ell_2/N} \psi^{N^2|s|-((Nr+\ell_1)+kN|s|),N^2|s|}_{T^2}(z) \notag \\
&= e^{2\pi i\ell_2/N} e^{-2\pi i(N-k-1)(N-\ell_2)/N} \psi^{(N(|s|-r-1)+N-\ell_1)+(N-k-1)N|s|),N^2s}_{T^2}(z) \notag \\
&= e^{2\pi i\ell_2/N} e^{-2\pi ik'\ell'_2/N} \psi^{(Nr'+\ell'_1)+k'N|s|),N^2|s|}_{T^2}(z). \notag
\end{align}},
\begin{align}
\Psi_{T^2/\mathbb{Z}^{(\ell_1,\ell_2)}_N}^{r,|s|}(-z,\tau) = e^{2\pi i\ell_2/N} \Psi_{T^2/\mathbb{Z}^{(N-\ell_1,N-\ell_2)}_N}^{|s|-r-1,|s|}(z,\tau). \label{-zshiftT2ZN}
\end{align}
The other relations except for Eq.~(\ref{ZshiftT2ZN}) are the same as the $T^2$ basis, where we note that the representation of $T^N$ transformation is diagonalized.
However, the $\mathbb{Z}_N$-shift invariant modes on the full $T^2/\mathbb{Z}_N$ shifted orbifold, i.e.~$(\ell_1,\ell_2)=(0,0)$, in particular, correspond to the modes on the $T^2/N \simeq T^{2'}$ with magnetic flux $s=M/N^2 \in 2\mathbb{Z}$.
In other words, the $\mathbb{Z}_N$-shift invariant modes behave like modular forms for $\widetilde{\Gamma}(2|M|/N^2)$, while the other modes correspond to modular forms for $\widetilde{\Gamma}(2|M|)$.


\section{Modular symmetry in magnetized $T^2/\mathbb{Z}_N$ twisted and shifted orbifold models}
\label{T2Z2twistshift}

In this section, we study the modular symmetry for the wavefunctions on the magnetized $T^2/\mathbb{Z}_N$ twisted and shifted orbifolds.
The modular symmetry remains on the $T^2/\mathbb{Z}_2$ twisted orbifold.
In order for the $T^2/\mathbb{Z}_2$ twisted orbifold to be consistent with the full $T^2/\mathbb{Z}_N$ shifted orbifold,
the following condition should be also satisfied,
\begin{align}
N-\ell_{1,2} \equiv \ell_{1,2}\ ({\rm mod}\ N),
\end{align}
for $^{\forall} \ell_{1,2} \in \mathbb{Z}_N$.
Therefore, the only full $T^2/\mathbb{Z}_2$ shifted orbifold is consistent with the $T^2/\mathbb{Z}_2$ twisted orbifold\footnote{The other $T^2/\mathbb{Z}_N$ twisted orbifolds with $N=3,4,6$, on the other hand, are not consistent with any full $T^2/\mathbb{Z}_N$ shifted orbifolds since they require $\ell_1=\ell_2$.}.
The wavefunctions on the magnetized $T^2/\mathbb{Z}_2$ twisted and shifted orbifold are expressed as
\begin{align}
&\Psi_{T^2/\mathbb{Z}^{(m;\ell_1,\ell_2)}_2}^{r,|s|}
= {\cal N}_2^{st}  \left( \Psi_{T^2/\mathbb{Z}^{(\ell_1,\ell_2)}_2}^{r,|s|} + (-1)^{m+\ell_2} \Psi_{T^2/\mathbb{Z}^{(\ell_1,\ell_2)}_2}^{|s|-r-\ell_1,|s|} \right)
= {\cal N}_2^{st} \left( \psi_{T^2/\mathbb{Z}_2^{m}}^{2r+\ell_1,4|s|} + (-1)^{\ell_2} \psi_{T^2/\mathbb{Z}_2^{m}}^{2r+\ell_1+2|s|,4|s|} \right) \notag \\
&= {\cal N}_2^{st} \left( \psi_{T^2}^{2r+\ell_1,4|s|} + (-1)^{\ell_2+m} \psi_{T^2}^{2(|s|-r-\ell_1)+\ell_1,4|s|} + (-1)^{\ell_2} \psi_{T^2}^{2(|s|+r)+\ell_1,4|s|} + (-1)^{m} \psi_{T^2}^{2(2|s|-r-\ell_1)+\ell_1,4|s|} \right), \notag \\
&\hspace{5.0cm} s \in 2\mathbb{Z}, \ r \in \mathbb{Z}_{\frac{|s|}{2}+1-\ell_1}, \ m, \ell_1, \ell_2 \in \mathbb{Z}_2, \label{twistshiftT2Z2exp}
\end{align}
where ${\cal N}_2^{st}$ is the normalization factor determined by remaining Eq.~(\ref{normalize}).
Note that $\ell^{(1,1)} \equiv \ell_1+\ell_2\ ({\rm mod}\ 2)$.
The numbers of the degenerate modes for $(m;\ell_1,\ell_2)=(0;0,0),(1;0,0)$ are $(|M|/8+1), (|M|/8-1)$, respectively, while the numbers of the degenerate modes for the other $(m;\ell_1,\ell_2)$ are $|M|/8$, where $M \in 8\mathbb{Z}$.
On this $T^2/\mathbb{Z}_2$ twisted and shifted orbifold basis, Eqs.~(\ref{rhoshiftZNS}) and (\ref{rhoshiftZNT}) 
as well as Eq.~(\ref{rhoSandT}) are deformed as
\begin{align}
\rho_{T^2/\mathbb{Z}_2^{(0;\ell_1,\ell_2)}}(\widetilde{S})_{rr',(\ell_1,\ell_2)(\ell'_1,\ell'_2)} &= e^{i\pi/4} \frac{2}{\sqrt{|s|}}  \cos \left( 2\pi (\ell_1/2+r) (\ell'_1/2+r')/|s| \right) \delta_{\ell_2,\ell'_1} \delta_{\ell_1,\ell'_2}, \label{rho+shiftZ2S} \\
\rho_{T^2/\mathbb{Z}_2^{(0;\ell_1,\ell_2)}}(\widetilde{T})_{rr',(\ell_1,\ell_2)(\ell'_1,\ell'_2)} &= e^{i\pi \left(\frac{\ell_1}{N}+r\right)^2/|s|} \delta_{r,r'} \delta_{\ell_1,\ell'_1} \delta_{\ell_2-\ell_1,\ell'_2}, \label{rho+shoftZ2T} \\
\rho_{T^2/\mathbb{Z}_2^{(1;\ell_1,\ell_2)}}(\widetilde{S})_{rr',(\ell_1,\ell_2)(\ell'_1,\ell'_2)} &= e^{i\pi/4} \frac{2i}{\sqrt{|s|}}  \sin \left( 2\pi (\ell_1/2+r) (\ell'_1/2+r')/|s| \right) \delta_{\ell_2,\ell'_1} \delta_{\ell_1,\ell'_2}, \label{rho-shiftZ2S} \\
\rho_{T^2/\mathbb{Z}_2^{(1;\ell_1,\ell_2)}}(\widetilde{T})_{rr',(\ell_1,\ell_2)(\ell'_1,\ell'_2)} &= e^{i\pi \left(\frac{\ell_1}{N}+r\right)^2/|s|} \delta_{r,r'} \delta_{\ell_1,\ell'_1} \delta_{\ell_2-\ell_1,\ell'_2}. \label{rho-shoftZ2T}
\end{align}
They satisfy the same relations as the $T^2/\mathbb{Z}_2$ twisted orbifold basis.
In particular, the $\mathbb{Z}_2$-shift invariant modes, i.e.~$(m;\ell_1,\ell_2)=(m;0,0)$, correspond to the modes on the $(T^2/N)/\mathbb{Z}_2 \simeq T^{2'}/\mathbb{Z}_2$ twisted orbifold with magnetic flux $s=M/4 \in 2\mathbb{Z}$.
For example, when $M=8\ (s=2)$, the $\mathbb{Z}_2$-shift invariant wavefunctions on the $T^2/\mathbb{Z}_2$ twisted and shifted orbifold basis are expressed as
\begin{align}
\begin{pmatrix}
\Psi_{T^2/\mathbb{Z}^{(0;0,0)}_2}^{0,2} \\ \Psi_{T^2/\mathbb{Z}^{(0;0,0)}_2}^{1,2}
\end{pmatrix}
&=
\begin{pmatrix}
\frac{1}{\sqrt{2}} \left( \psi^{0,8}_{T^2}+\psi^{4,8}_{T^2} \right) \\ \frac{1}{\sqrt{2}} \left( \psi^{2,8}_{T^2}+\psi^{6,8}_{T^2} \right)
\end{pmatrix}, \label{psishifttwistM8}
\end{align}
and the $S$ and $T$ transformations for Eq.~(\ref{psishifttwistM8}) are the same as Eq.~(\ref{M2rep}).
When $M=16\ (s=4)$, the $\mathbb{Z}_2$-shift invariant wavefunctions on the $T^2/\mathbb{Z}_2$ twisted and shifted orbifold basis are expressed as
\begin{align}
\begin{pmatrix}
\Psi_{T^2/\mathbb{Z}^{(0;0,0)}_2}^{0,4} \\ \Psi_{T^2/\mathbb{Z}^{(0;0,0)}_2}^{1,4} \\ \Psi_{T^2/\mathbb{Z}^{(0;0,0)}_2}^{2,4}
\end{pmatrix}
&=
\begin{pmatrix}
\frac{1}{\sqrt{2}} \left( \psi^{0,16}_{T^2}+\psi^{8,16}_{T^2} \right) \\ \frac{1}{2} \left( \psi^{2,16}_{T^2}+\psi^{6,16}_{T^2}+\psi^{10,16}_{T^2}+\psi^{14,16}_{T^2} \right) \\ \frac{1}{\sqrt{2}} \left( \psi^{4,16}_{T^2}+\psi^{12,16}_{T^2} \right)
\end{pmatrix}, \label{psishifttwistM16m0l0} \\
\Psi_{T^2/\mathbb{Z}^{(0;0,0)}_2}^{1,4}
&=
\frac{1}{2} \left( \psi^{2,16}_{T^2}-\psi^{6,16}_{T^2}+\psi^{10,16}_{T^2}-\psi^{14,16}_{T^2} \right),
\label{psishifttwistM16m1l0}
\end{align}
and the representations of the $S$ and $T$ transformations are the same as Eqs.~(\ref{M4evenrep}) and (\ref{M4oddrep}).
We express the all wavefunctions on the $T^2/\mathbb{Z}_2$ twisted and shifted orbifold base for $M=8,16$ and the representations of the $S$ and $T$ transformations for them in Appendix~\ref{exM816}.

As a result, when $M=0$ (mod 8), both the $\mathbb{Z}_2$ twist and the full $\mathbb{Z}_2$ shift are 
consistent with the modular symmetry.
The wavefunctions can be decomposed into smaller representations by their eigenvalues.
Thus, a combination between the $\mathbb{Z}_2$ twist and the full $\mathbb{Z}_2$ shift provides us 
with a reduction of reducible representations towards irreducible representations 
$\widetilde{\Gamma}'_{2|M|} \equiv \widetilde{\Gamma}/\widetilde{\Gamma}(2|M|)$.

\section{Conclusion}
\label{conclusion}
 
We have studied the modular symmetry of wavefunctions on the magnetized $T^2 \simeq \mathbb{C}/\Lambda$.
When the magnetic flux $M$ is even, 
the wavefunctions behave as modular forms of weight $1/2$ and 
represent the double covering group of $\Gamma \equiv SL(2,\mathbb{Z})$, 
$\widetilde{\Gamma} \equiv \widetilde{SL}(2,\mathbb{Z})$.
Each wavefunction on $T^2$ with the magnetic flux $M$ transforms under
 $\widetilde{\Gamma}(2|M|)$.
Then, $|M|$ zero-modes as well as massive modes are representations of  the quotient group $\widetilde{\Gamma}'_{2|M|} \equiv \widetilde{\Gamma}/\widetilde{\Gamma}(2|M|)$.

If we change the Wilson line  $\zeta \rightarrow \zeta+1$ simultaneously with the $T$ transformation of the modular transformations, $T^2$ with any magnetic flux $M$ is consistent with the modular transformations.
However, the zero-mode wavefunctions after the $T$ transformation are expanded by the all excited-mode wavefunctions before the $T$ transformation.


We have also studied the modular symmetry for the wavefunctions on various magnetized $T^2/\mathbb{Z}_N$ orbifolds.
The $T^2/\mathbb{Z}_N$ twisted orbifold can be constructed for $N=2,3,4,6$.
However, the modulus $\tau \equiv e_2/e_1$ is fixed as $\tau=e^{2\pi i/N}$ for $N=3,4,6$ while any $\tau$ is allowed for $N=2$.
It means that the only $ST, S, ST^{-1}$ transformations of the modular transformations remain for $N=3,4,6$, respectively.
They correspond to $Z_{2N}$ symmetries.
On the other hand, there remains the full modular symmetry for $N=2$.
The representations of the modular transformations on the $T^2/\mathbb{Z}_2$ twisted orbifold basis satisfy the same algebraic relations as the representations on the $T^2$ basis.
However,  the representations on the $T^2$ basis are decomposed into smaller representations on the $T^2/\mathbb{Z}_2$ twisted orbifold basis.

In order for the $T^2/\mathbb{Z}_N$ shifted orbifold to be consistent with the modular transformations, all $\mathbb{Z}_N$-shifted points should be identified, where we call it the full $T^2/\mathbb{Z}_N$ shifted orbifold.
The full $T^2/\mathbb{Z}_N$ shifted orbifold with the magnetic flux $M$ corresponds to $T^{2'} \simeq \mathbb{C}/\Lambda', \Lambda' \equiv \Lambda/N$ with the magnetic flux $s \equiv M/N^2 \in 2\mathbb{Z}$.
In particular, the $\mathbb{Z}_N$-shift invariant modes correspond to the modes on $T^2/N \simeq T^{2'}$.
Therefore, the $\mathbb{Z}_N$-shift invariant modes behave like modular forms for $\widetilde{\Gamma}(2|M|/N^2)$, while the other modes behave as modular forms for $\widetilde{\Gamma}(2|M|)$.

Furthermore, the only full $T^2/\mathbb{Z}_2$ shifted orbifold is consistent with the $T^2/\mathbb{Z}_2$ twisted orbifold.
On that $T^2/\mathbb{Z}_2$ twisted and shifted orbifold, the $\mathbb{Z}_2$-shift invariant modes correspond to the modes on the $(T^2/N)/\mathbb{Z}_2$ twisted orbifold with the magnetic flux $s \equiv M/4 \in 2\mathbb{Z}$.

The wavefunctions on $T^2$ are decomposed into smaller representations 
by the $\mathbb{Z}_2$ twist and shift.
They provide us with a reduction of representations towards irreducible representations.
Also, the combination of the $\mathbb{Z}_2$ twist and shift provides us with a new approach to realize three generations 
from the phenomenological viewpoints \footnote{See \cite{Abe:2008sx,Abe:2015yva} for classifications of three-generation models by 
$T^2/\mathbb{Z}_N$ twisted orbifolds.}.
It is interesting to study three-generation models by a combination of $\mathbb{Z}_2$ twist and shift.
We would study elsewhere.

\vspace{1.5 cm}
\noindent
{\large\bf Acknowledgement}\\

The authors would like to thank Y. Ogawa for useful comments.
T. K. was supported in part by MEXT KAKENHI Grant Number JP19H04605. 
H. U. was supported by Grant-in-Aid for JSPS Research Fellows No. 20J20388.

\appendix
\section*{Appendix}


\section{Extension for generalized $CP$ symmetry with the modular symmetry on the magnetized $T^2$}
\label{gCP}

Here, we study the extension for generalized $CP$ symmetry with the modular transformations on the magnetized $T^2$~\footnote{
See \cite{Baur:2019kwi,Novichkov:2019sqv} for the relation between the modular symmetry and $CP$ symmetry.
See also Ref.~\cite{Kobayashi:2020uaj} and references therein for $CP$ in superstring theory.}.
The $CP$ transformation for the modulus $\tau$ is defined as $CP: \tau \rightarrow -\bar{\tau}$, where it remains ${\rm Im}(-\bar{\tau})>0$.
It is derived from
\begin{align}
&CP:
\begin{pmatrix}
e_2 \\ e_1
\end{pmatrix}
\rightarrow
\begin{pmatrix}
e_2^{CP} \\ e_1^{CP}
\end{pmatrix}
=
\begin{pmatrix}
1 & 0 \\
0 & -1
\end{pmatrix}
\begin{pmatrix}
\bar{e}_2 \\ \bar{e}_1
\end{pmatrix}, \quad
CP =
\begin{pmatrix}
1 & 0 \\
0 & -1
\end{pmatrix},
\label{CP} \\
&CP: z \equiv \frac{u}{e_1} \rightarrow z^{CP} \equiv \frac{u^{CP}}{e_1^{CP}} = \frac{\bar{u}}{-\bar{e}_1} = -\bar{z}, \label{CPz} \\
&CP: \tau \equiv \frac{e_2}{e_1} \rightarrow \tau^{CP} \equiv \frac{e_2^{CP}}{e_1^{CP}} = \frac{\bar{e}_2}{-\bar{e}_1} = -\bar{\tau}. \label{CPtau}
\end{align}
The $CP$ matrix in Eq.~(\ref{CP}) satisfies the following relations,
\begin{align}
CP^2=\mathbb{I}, \quad (CP)S(CP)^{-1}=S^{-1}, \quad (CP)T(CP)^{-1}=T^{-1}. \label{CPST}
\end{align}
When we also consider the above $CP$ transformation in addition to the modular transformations, the modular group $\Gamma=SL(2,\mathbb{Z})$ is extended to $\Gamma^{\ast} \equiv SL(2,\mathbb{Z}) \rtimes \mathbb{Z}_2^{CP} \simeq GL(2,\mathbb{Z})$.
Under the extended modular transformation by $\gamma^{\ast}=
\begin{pmatrix}
a & b \\
c & d
\end{pmatrix} \in \Gamma^{\ast}$, $(z,\tau)$ transforms as
\begin{align}
\gamma^{\ast}: (z,\tau) \rightarrow
\left\{
\begin{array}{cc}
\left( \frac{z}{c\tau+d}, \frac{a\tau+b}{c\tau+d} \right), & ({\rm det} \gamma^{\ast} = 1) \\
\left( \frac{\bar{z}}{c\bar{\tau}+d}, \frac{a\bar{\tau}+b}{c\bar{\tau}+d} \right), & ({\rm det} \gamma^{\ast} = -1)
\end{array}
\right. ,
\label{gammaast}
\end{align}
where the above in Eq.~(\ref{gammaast}) is just modular transformation and the below in Eq.~(\ref{gammaast}) contains odd numbers of $CP$ transformation.
It leads to redefine the automorphy factor as
\begin{align}
J_k(\gamma^{\ast}, \tau) \equiv
\left\{
\begin{array}{cc}
(c\tau+d)^k, & ({\rm det} \gamma^{\ast} = 1) \\
(c\bar{\tau}+d)^k, & ({\rm det} \gamma^{\ast} = -1)
\end{array}
\right. , \quad
\gamma^{\ast} =
\begin{pmatrix}
a & b \\
c & d 
\end{pmatrix}
\in \Gamma^{\ast}, \label{reautomorphyfact}
\end{align}
where it satisfies Eq.~(\ref{J}).
Note that $\gamma^{\ast}$ does not mean the complex conjugate of $\gamma$ but an element of $\Gamma^{\ast}$.

In order to see how the wavefunctions on the magnetized $T^2$ transform under the extended modular transformation by $\gamma^{\ast} \in \Gamma^{\ast}$, furthermore, we consider the double covering group of $\Gamma^{\ast} \simeq GL(2,\mathbb{Z})$, $\widetilde{\Gamma^{\ast}} \simeq \widetilde{GL}(2,\mathbb{Z})$, similar to Eq.~(\ref{Gammatilde}).
Note that only Eq.~(\ref{Aalphabeta}) is redefined as
\begin{align}
A(\gamma_1^{\ast},\gamma_2^{\ast}) =
\left( {\rm det} \gamma_1^{\ast}, {\rm det} \gamma_2^{\ast} \right)_H \left( \frac{\chi(\gamma_1^{\ast} \gamma_2^{\ast})}{\chi(\gamma_1^{\ast})}, \frac{\chi(\gamma_1^{\ast} \gamma_2^{\ast})}{\chi(\gamma_2^{\ast}) {\rm det} \gamma_1^{\ast}} \right)_H. \label{Aalphabetaast}
\end{align}
(See Ref.~\cite{Duncan:2018wbw}.)
In particular, we set
\begin{align}
\widetilde{CP} \equiv [CP,1]. \label{CPtilde}
\end{align}
Then, Eqs.~(\ref{CPtilde}), (\ref{IST}), and (\ref{inverseSTZ}) lead to the following relations,
\begin{align}
(\widetilde{CP})^2=[\mathbb{I},-1]=\widetilde{Z}^2, \quad (\widetilde{CP})^4=[\mathbb{I},1]=\mathbb{I}=\widetilde{Z}^4, \quad (\widetilde{CP})^{-1}=[(CP)^{-1},-1],
\notag \\
(\widetilde{CP}) \widetilde{S} (\widetilde{CP})^{-1} = [S^{-1},1] = \widetilde{S}^{-1}, \quad (\widetilde{CP}) \widetilde{T} (\widetilde{CP})^{-1} = [T^{-1},1] = \widetilde{T}^{-1}, \label{tildeCPrel}
\end{align}
in addition to Eq.~(\ref{tildealgebrarel}).
The automorphy factor is the same as Eq.~(\ref{Jtilde}) and satisfies Eq.~(\ref{Jtilderel}), where we should apply Eqs.~(\ref{reautomorphyfact}) and (\ref{Aalphabetaast}).

Here, we study the $CP$ transformation of the fields on the magnetized $T^2$.
In addition to Eqs.~(\ref{CPz}) and (\ref{CPtau}), it is also needed that the magnetic flux $M$ is flipped as
\begin{align}
CP: M \rightarrow -M. \label{fluxCP}
\end{align}
In this case, any field in section~\ref{wavT2} after the $CP$ transformation corresponds to the complex conjugate of the field.
In particular, the wavefunctions of the $n$-th excited-modes in Eq.~(\ref{psin}), including the zero-modes in Eq.~(\ref{psizero}),  transform as
\begin{align}
&CP: \psi_n^{j,M}(z, \tau) \rightarrow \psi_n^{j,|M|}(-\bar{z}, -\bar{\tau}) = \overline{\psi_n^{j,|M|}(z, \tau)}, \label{psiCP}
\end{align}
under the $CP$ transformation.
Considering Eqs.~(\ref{wavemodularform}) and (\ref{psiCP}), we can obtain the following form,
\begin{align}
\psi_n^{j,|M|}(\widetilde{CP}(z,\tau)) &= \widetilde{J}_{1/2}(\widetilde{CP}, \tau) \sum_{k=0}^{|M|-1} \rho(\widetilde{CP})_{jk} \overline{\psi_n^{k,|M|}(z,\tau)}, \label{waveCP} \\
\quad \widetilde{J}_{1/2}(\widetilde{CP}, \tau) =& (-1)^{1/2} = e^{-i\pi /2} = -i, \quad \rho(\widetilde{CP})_{jk} = i \delta_{j,k}.
\label{JrhoCP}
\end{align}
We can also check the following relations,
\begin{align}
\widetilde{J}_{1/2}((\widetilde{CP})^{-1}, \tau) =& -(-1)^{1/2} = -e^{-i\pi /2} = i, \quad \rho(\widetilde{CP})^{-1}_{jk} = -i \delta_{j,k}. \label{JrhoCP-1}
\end{align} 
From Eqs.~(\ref{JrhoCP}), (\ref{JrhoCP-1}), and (\ref{rhoSandT}), we can obtain the following relations,
\begin{align}
\widetilde{J}_{1/2}(\widetilde{Z}^2,\tau)=\widetilde{J}_{1/2}((\widetilde{CP})^2,\tau)=-1, &\quad \rho(\widetilde{Z})^2=\rho(CP)^2=-\delta_{j,k}, \label{CP2} \\
\widetilde{J}_{1/2}(\widetilde{Z}^4,\tau)=\widetilde{J}_{1/2}((\widetilde{CP})^4,\tau)=-1, &\quad \rho(\widetilde{Z})^4=\rho(CP)^4=\delta_{j,k}, \label{CP2} \\
\widetilde{J}_{1/2}((\widetilde{CP})\widetilde{S}(\widetilde{CP})^{-1},\tau)=\widetilde{J}_{1/2}(\widetilde{S}^{-1},\tau), &\quad [\rho(\widetilde{CP})\overline{\rho(\widetilde{S})}\rho(\widetilde{CP})^{-1}]_{jk}=\rho(\widetilde{S})^{-1}_{jk}, \label{CPSCP-1} \\
\widetilde{J}_{1/2}((\widetilde{CP})\widetilde{T}(\widetilde{CP})^{-1},\tau)=\widetilde{J}_{1/2}(\widetilde{T}^{-1},\tau), &\quad [\rho(\widetilde{CP})\overline{\rho(\widetilde{T})}\rho(\widetilde{CP})^{-1}]_{jk}=\rho(\widetilde{T})^{-1}_{jk}, \label{CPTCP-1}
\end{align}
which are the representations of Eq.~(\ref{tildeCPrel}).
Then, $\rho$ becomes the representation of $\widetilde{\Gamma^{\ast}}'_{2|M|} \equiv \widetilde{\Gamma^{\ast}}/\widetilde{\Gamma}(2|M|)$.


\section{Modular transformation with gauge transformation}
\label{withgauge}

Here, we derive Eq.~(\ref{Twilsonzeromassive}),
\begin{align}
&T: \psi_0^{j,|M|}(z+\zeta, \tau) \rightarrow \psi_0^{j,|M|}(z+\zeta+1, \tau+1) \notag \\
&= \left( \frac{|M|}{{\cal A}^2} \right)^{1/4} e^{i\pi |M|(z+\zeta+1)\frac{{\rm Im}(z+\zeta)}{{\rm Im}\tau}} \sum_{l \in \mathbf{Z}} e^{i\pi |M|(\tau+1) \left( \frac{j}{|M|}+l \right)^2} e^{2\pi i|M| (z+\zeta) \left( \frac{j}{|M|}+l \right)} \notag \\
&= (-1)^2 e^{i\pi \frac{j^2}{|M|}} \left( \frac{|M|}{{\cal A}^2} \right)^{1/4} e^{i\pi |M|(z+\zeta)\frac{{\rm Im}(z+\zeta)}{{\rm Im}\tau}} \sum_{l \in \mathbf{Z}} e^{i\pi |M| \left(\frac{{\rm Im}(z+\zeta)}{{\rm Im}\tau}+\frac{j}{|M|}+l\right)} e^{i\pi |M|\tau \left( \frac{j}{|M|}+l \right)^2} e^{2\pi i|M| (z+\zeta) \left( \frac{j}{|M|}+l \right)} \notag \\
&= (-1)^2 e^{i\pi \frac{j^2}{|M|}} e^{-\frac{\pi |M|}{8{\rm Im}\tau}} \sum_{n=0}^{\infty} \frac{1}{n!} \left(i\sqrt{\frac{\pi |M|}{8{\rm Im}\tau}}\right)^n \left( \frac{|M|}{{\cal A}^2} \right)^{1/4} \notag \\
&\times e^{i\pi |M| (z+\zeta) \frac{{\rm Im}(z+\zeta)}{{\rm Im}\tau}} \sum_{l \in \mathbf{Z}} e^{i\pi |M|\tau \left( \frac{j}{|M|}+l \right)^2} e^{2\pi i|M| (z+\zeta) \left( \frac{j}{|M|}+l \right)} H_{n} \left( \sqrt{2\pi |M|{\rm Im}\tau} \left( \frac{{\rm Im}z}{{\rm Im}\tau}+\frac{j}{|M|}+l \right) \right) \notag \\
&= (-1)^j e^{i\pi \frac{j^2}{|M|}} e^{-\frac{\pi |M|}{8{\rm Im}\tau}} \sum_{n=0}^{\infty} \frac{1}{\sqrt{n!}} \left(i\sqrt{\frac{\pi |M|}{4{\rm Im}\tau}}\right)^n \psi_{n}^{j,|M|}(z+\zeta,\tau), \label{Twilsonzeromassivecal}
\end{align}
where we use the following the generating function of the Hermite function,
\begin{align}
e^{-y^2+2xy} = \sum_{n=0} H_n(x) \frac{y^n}{n!}. \label{Hermitobokansu}
\end{align}


\section{Examples of the $T^2/\mathbb{Z}_2$ twisted and shifted orbifold base}
\label{exM816}

Here, we express examples of the wavefunctions on the $T^2/\mathbb{Z}_2$ twisted and shifted orbifold basis and the representations of the $S$ and $T$ transformations for them.
In particular, we show them for $M=8$ and $16$.

When $M=8\ (s=2)$, the wavefunctions on the $T^2/\mathbb{Z}_2$ twisted and shifted orbifold basis are expressed as
\begin{align}
\begin{pmatrix}
\Psi_{T^2/\mathbb{Z}^{(0;0,0)}_2}^{0,2} \\ \Psi_{T^2/\mathbb{Z}^{(0;0,0)}_2}^{1,2}
\end{pmatrix}
&=
\begin{pmatrix}
\frac{1}{\sqrt{2}} \left( \psi^{0,8}_{T^2}+\psi^{4,8}_{T^2} \right) \\ \frac{1}{\sqrt{2}} \left( \psi^{2,8}_{T^2}+\psi^{6,8}_{T^2} \right)
\end{pmatrix}, \label{psishifttwistM8m0l0} \\
\begin{pmatrix}
\Psi_{T^2/\mathbb{Z}^{(0;0,1)}_2}^{0,2} \\ \Psi_{T^2/\mathbb{Z}^{(0;1,0)}_2}^{0,2} \\ \Psi_{T^2/\mathbb{Z}^{(0;1,1)}_2}^{0,2}
\end{pmatrix}
&=
\begin{pmatrix}
\frac{1}{\sqrt{2}} \left( \psi^{0,8}_{T^2}-\psi^{4,8}_{T^2} \right) \\ \frac{1}{2} \left( \psi^{1,8}_{T^2}+\psi^{3,8}_{T^2}+\psi^{5,8}_{T^2}+\psi^{7,8}_{T^2} \right) \\ \frac{1}{2} \left( \psi^{1,8}_{T^2}-\psi^{3,8}_{T^2}-\psi^{5,8}_{T^2}+\psi^{7,8}_{T^2} \right)
\end{pmatrix}, \label{psishifttwistM8m0} \\ 
\begin{pmatrix}
\Psi_{T^2/\mathbb{Z}^{(1;0,1)}_2}^{0,2} \\ \Psi_{T^2/\mathbb{Z}^{(0;1,0)}_2}^{0,2} \\ \Psi_{T^2/\mathbb{Z}^{(0;1,1)}_2}^{0,2}
\end{pmatrix}
&=
\begin{pmatrix}
\frac{1}{\sqrt{2}} \left( \psi^{2,8}_{T^2}-\psi^{6,8}_{T^2} \right) \\ \frac{1}{2} \left( \psi^{1,8}_{T^2}-\psi^{3,8}_{T^2}+\psi^{5,8}_{T^2}-\psi^{7,8}_{T^2} \right) \\ \frac{1}{2} \left( \psi^{1,8}_{T^2}+\psi^{3,8}_{T^2}-\psi^{5,8}_{T^2}-\psi^{7,8}_{T^2} \right)
\end{pmatrix}. \label{psishifttwistM8m1}
\end{align}
The representations of the $S$ and $T$ transformations for Eq.~(\ref{psishifttwistM8m0l0}) are expressed as
\begin{align}
\rho_{T^2/\mathbb{Z}^{(0;0,0)}_2}(\widetilde{S}) = \frac{e^{i\pi /4}}{\sqrt{2}}
\begin{pmatrix}
1 & 1 \\
1 & -1
\end{pmatrix}, \quad
\rho_{T^2/\mathbb{Z}^{(0;0,0)}_2}(\widetilde{T}) =
\begin{pmatrix}
1 & 0 \\
0 & i
\end{pmatrix},
\label{M8m0l0}
\end{align}
which are the same as Eq.~(\ref{M2rep}).
The representations of the $S$ and $T$ transformations for Eqs.~(\ref{psishifttwistM8m0}) and (\ref{psishifttwistM8m1}) are expressed as
\begin{align}
\rho_{T^2/\mathbb{Z}^{(0;\ell_1,\ell_2)}_2}(\widetilde{S}) = e^{i\pi /4}
\begin{pmatrix}
0 & 1 & 0 \\
1 & 0 & 0 \\
0 & 0 & 1
\end{pmatrix}, \quad
\rho_{T^2/\mathbb{Z}^{(0;\ell_1,\ell_2)}_2}(\widetilde{T}) =
\begin{pmatrix}
1 & 0 & 0 \\
0 & 0 & e^{i\pi /8} \\
0 & e^{i\pi /8} & 0
\end{pmatrix},
\label{M8m0l}
\end{align}
and
\begin{align}
\rho_{T^2/\mathbb{Z}^{(1;\ell_1,\ell_2)}_2}(\widetilde{S}) = e^{i\pi /4} i 
\begin{pmatrix}
0 & 1 & 0 \\
1 & 0 & 0 \\
0 & 0 & 1
\end{pmatrix}, \quad
\rho_{T^2/\mathbb{Z}^{(1;\ell_1,\ell_2)}_2}(\widetilde{T}) =
\begin{pmatrix}
i & 0 & 0 \\
0 & 0 & e^{i\pi /8} \\
0 & e^{i\pi /8} & 0
\end{pmatrix},
\label{M8m0l}
\end{align}
respectively.

When $M=16\ (s=4)$, the wavefunctions on the $T^2/\mathbb{Z}_2$ twisted and shifted orbifold basis are expressed as
\begin{align}
\begin{pmatrix}
\Psi_{T^2/\mathbb{Z}^{(0;0,0)}_2}^{0,4} \\ \Psi_{T^2/\mathbb{Z}^{(0;0,0)}_2}^{1,4} \\ \Psi_{T^2/\mathbb{Z}^{(0;0,0)}_2}^{2,4}
\end{pmatrix}
&=
\begin{pmatrix}
\frac{1}{\sqrt{2}} \left( \psi^{0,16}_{T^2}+\psi^{8,16}_{T^2} \right) \\ \frac{1}{2} \left( \psi^{2,16}_{T^2}+\psi^{6,16}_{T^2}+\psi^{10,16}_{T^2}+\psi^{14,16}_{T^2} \right) \\ \frac{1}{\sqrt{2}} \left( \psi^{4,16}_{T^2}+\psi^{12,16}_{T^2} \right)
\end{pmatrix}, \label{psishifttwistM16m0l0} \\
\begin{pmatrix}
\Psi_{T^2/\mathbb{Z}^{(0;0,1)}_2}^{0,4} \\
\Psi_{T^2/\mathbb{Z}^{(0;0,1)}_2}^{1,4} \\
\Psi_{T^2/\mathbb{Z}^{(0;1,0)}_2}^{0,4} \\
\Psi_{T^2/\mathbb{Z}^{(0;1,0)}_2}^{1,4} \\
\Psi_{T^2/\mathbb{Z}^{(0;1,1)}_2}^{0,4} \\
\Psi_{T^2/\mathbb{Z}^{(0;1,1)}_2}^{1,4}
\end{pmatrix}
&=
\begin{pmatrix}
\frac{1}{\sqrt{2}} \left( \psi^{0,16}_{T^2}-\psi^{8,16}_{T^2} \right) \\
\frac{1}{2} \left( \psi^{2,16}_{T^2}-\psi^{6,16}_{T^2}-\psi^{10,16}_{T^2}+\psi^{14,16}_{T^2} \right) \\
\frac{1}{2} \left( \psi^{1,16}_{T^2}+\psi^{7,16}_{T^2}+\psi^{9,16}_{T^2}+\psi^{15,16}_{T^2} \right) \\
\frac{1}{2} \left( \psi^{3,16}_{T^2}+\psi^{5,16}_{T^2}+\psi^{11,16}_{T^2}+\psi^{13,16}_{T^2} \right) \\
\frac{1}{2} \left( \psi^{1,16}_{T^2}-\psi^{7,16}_{T^2}-\psi^{9,16}_{T^2}+\psi^{15,16}_{T^2} \right) \\
\frac{1}{2} \left( \psi^{3,16}_{T^2}-\psi^{5,16}_{T^2}-\psi^{11,16}_{T^2}+\psi^{13,16}_{T^2} \right) \\
\end{pmatrix}, \label{psishifttwistM16m0} \\
\Psi_{T^2/\mathbb{Z}^{(0;0,0)}_2}^{1,4}
&=
\frac{1}{2} \left( \psi^{2,16}_{T^2}-\psi^{6,16}_{T^2}+\psi^{10,16}_{T^2}-\psi^{14,16}_{T^2} \right),
\label{psishifttwistM16m1l0} \\
\begin{pmatrix}
\Psi_{T^2/\mathbb{Z}^{(1;0,1)}_2}^{0,4} \\
\Psi_{T^2/\mathbb{Z}^{(1;0,1)}_2}^{1,4} \\
\Psi_{T^2/\mathbb{Z}^{(1;1,0)}_2}^{0,4} \\
\Psi_{T^2/\mathbb{Z}^{(1;1,0)}_2}^{1,4} \\
\Psi_{T^2/\mathbb{Z}^{(1;1,1)}_2}^{0,4} \\
\Psi_{T^2/\mathbb{Z}^{(1;1,1)}_2}^{1,4}
\end{pmatrix}
&=
\begin{pmatrix}
\frac{1}{2} \left( \psi^{2,16}_{T^2}+\psi^{6,16}_{T^2}-\psi^{10,16}_{T^2}-\psi^{14,16}_{T^2} \right) \\
\frac{1}{\sqrt{2}} \left( \psi^{4,16}_{T^2}-\psi^{12,16}_{T^2} \right) \\
\frac{1}{2} \left( \psi^{1,16}_{T^2}-\psi^{7,16}_{T^2}+\psi^{9,16}_{T^2}-\psi^{15,16}_{T^2} \right) \\
\frac{1}{2} \left( \psi^{3,16}_{T^2}-\psi^{5,16}_{T^2}+\psi^{11,16}_{T^2}-\psi^{13,16}_{T^2} \right) \\
\frac{1}{2} \left( \psi^{1,16}_{T^2}+\psi^{7,16}_{T^2}-\psi^{9,16}_{T^2}-\psi^{15,16}_{T^2} \right) \\
\frac{1}{2} \left( \psi^{3,16}_{T^2}+\psi^{5,16}_{T^2}-\psi^{11,16}_{T^2}-\psi^{13,16}_{T^2} \right) \\
\end{pmatrix}. \label{psishifttwistM16m1}
\end{align}
The representations of the $S$ and $T$ transformations for Eqs.~(\ref{psishifttwistM16m0}) and (\ref{psishifttwistM16m1}) are expressed as
\begin{align}
\rho_{T^2/\mathbb{Z}^{(0;0,0)}_2}(\widetilde{S}) = \frac{e^{i\pi /4}}{2}
\begin{pmatrix}
1 & \sqrt{2} & 1 \\
\sqrt{2} & 0 & -\sqrt{2} \\
1 & -\sqrt{2} & 1
\end{pmatrix}, \quad
\rho_{T^2/\mathbb{Z}^{(0;0,0)}_2}(\widetilde{T}) =
\begin{pmatrix}
1 & 0 & 0 \\
0 & e^{i\pi /4} & 0 \\
0 & 0 & -1
\end{pmatrix},
\label{M16m0l0}
\end{align}
which are the same as Eq.~(\ref{M4evenrep}), and
\begin{align}
\rho_{T^2/\mathbb{Z}^{(0;0,0)}_2}(\widetilde{S}) = e^{3\pi i/4}, \quad \rho_{T^2/\mathbb{Z}^{(0;0,0)}_2}(\widetilde{T}) = e^{i\pi /4}, \label{M16m1l0}
\end{align}
which are the same as Eq.~(\ref{M4oddrep}), respectively. 
The representations of the $S$ and $T$ transformations for Eqs.~(\ref{psishifttwistM16m0}) and (\ref{psishifttwistM16m1}) are expressed as
\begin{align}
\rho_{T^2/\mathbb{Z}^{(0;\ell_1,\ell_2)}_2}(\widetilde{S}) &= e^{i\pi /4}
\begin{pmatrix}
0 & 0 & 1/\sqrt{2} & 1/\sqrt{2} & 0 & 0 \\
0 & 0 & 1/\sqrt{2} & -1/\sqrt{2} & 0 & 0 \\
1/\sqrt{2} & 1/\sqrt{2} & 0 & 0 & 0 & 0 \\
1/\sqrt{2} & -1/\sqrt{2} & 0 & 0 & 0 & 0 \\
0 & 0 & 0 & 0 & \cos (\pi /8) & \sin (\pi /8) \\
0 & 0 & 0 & 0 & \sin (\pi /8) & -\cos (\pi /8) \\
\end{pmatrix}, \notag \\
\rho_{T^2/\mathbb{Z}^{(0;\ell_1,\ell_2)}_2}(\widetilde{T}) &=
\begin{pmatrix}
1 & 0 & 0 & 0 & 0 & 0 \\
0 & e^{i\pi /4} & 0 & 0 & 0 & 0 \\
0 & 0 & 0 & 0 & e^{i\pi /16} & 0 \\
0 & 0 & 0 & 0 & 0 & e^{i\pi /16} i \\
0 & 0 & e^{i\pi /16} & 0 & 0 & 0 \\
0 & 0 & 0 & e^{i\pi /16} i & 0 & 0
\end{pmatrix},
\label{M8m0l}
\end{align}
and
\begin{align}
\rho_{T^2/\mathbb{Z}^{(1;\ell_1,\ell_2)}_2}(\widetilde{S}) &= e^{i\pi /4} i
\begin{pmatrix}
0 & 0 & 1/\sqrt{2} & 1/\sqrt{2} & 0 & 0 \\
0 & 0 & 1/\sqrt{2} & -1/\sqrt{2} & 0 & 0 \\
1/\sqrt{2} & 1/\sqrt{2} & 0 & 0 & 0 & 0 \\
1/\sqrt{2} & -1/\sqrt{2} & 0 & 0 & 0 & 0 \\
0 & 0 & 0 & 0 & \sin (\pi /8) & \cos (\pi /8) \\
0 & 0 & 0 & 0 & \cos (\pi /8) & -\sin (\pi /8) \\
\end{pmatrix}, \notag \\
\rho_{T^2/\mathbb{Z}^{(1;\ell_1,\ell_2)}_2}(\widetilde{T}) &=
\begin{pmatrix}
e^{i\pi /4} & 0 & 0 & 0 & 0 & 0 \\
0 &-1 & 0 & 0 & 0 & 0 \\
0 & 0 & 0 & 0 & e^{i\pi /16} & 0 \\
0 & 0 & 0 & 0 & 0 & e^{i\pi /16} i \\
0 & 0 & e^{i\pi /16} & 0 & 0 & 0 \\
0 & 0 & 0 & e^{i\pi /16} i & 0 & 0
\end{pmatrix},
\label{M8m0l}
\end{align}
respectively.



\begin{thebibliography}{99}






\bibitem{Altarelli:2010gt}
G.~Altarelli and F.~Feruglio,
Rev.\ Mod.\ Phys.\ {\bf 82} (2010) 2701
[arXiv:1002.0211 [hep-ph]].



\bibitem{Ishimori:2010au}
H.~Ishimori, T.~Kobayashi, H.~Ohki, Y.~Shimizu, H.~Okada and M.~Tanimoto,
Prog.\ Theor.\ Phys.\ Suppl.\ {\bf 183} (2010) 1
[arXiv:1003.3552 [hep-th]].



\bibitem{Ishimori:2012zz}
H.~Ishimori, T.~Kobayashi, H.~Ohki, H.~Okada, Y.~Shimizu and M.~Tanimoto,
Lect.\ Notes Phys.\ {\bf 858} (2012) 1, Springer.

\bibitem{Hernandez:2012ra}
D.~Hernandez and A.~Y.~Smirnov,
Phys.\ Rev.\ D {\bf 86} (2012) 053014
[arXiv:1204.0445 [hep-ph]].

\bibitem{King:2013eh}
S.~F.~King and C.~Luhn,
Rept.\ Prog.\ Phys.\ {\bf 76} (2013) 056201
[arXiv:1301.1340 [hep-ph]].

\bibitem{King:2014nza} 
S.~F.~King, A.~Merle, S.~Morisi, Y.~Shimizu and M.~Tanimoto,
New J.\ Phys.\ {\bf 16}, 045018 (2014)
[arXiv:1402.4271 [hep-ph]].


\bibitem{Tanimoto:2015nfa}
M.~Tanimoto,
AIP Conf.\ Proc.\ {\bf 1666} (2015) 120002.

\bibitem{King:2017guk}
S.~F.~King,
Prog.\ Part.\ Nucl.\ Phys.\ {\bf 94} (2017) 217
[arXiv:1701.04413 [hep-ph]].

\bibitem{Petcov:2017ggy}
S.~T.~Petcov,
Eur.\ Phys.\ J.\ C {\bf 78} (2018) no.9, 709
[arXiv:1711.10806 [hep-ph]].

 

\bibitem{Kobayashi:2006wq}
T.~Kobayashi, H.~P.~Nilles, F.~Ploger, S.~Raby and M.~Ratz,
Nucl. Phys. B \textbf{768}, 135-156 (2007)
[arXiv:hep-ph/0611020 [hep-ph]].

\bibitem{Abe:2009vi}
H.~Abe, K.~S.~Choi, T.~Kobayashi and H.~Ohki,
Nucl. Phys. B \textbf{820}, 317-333 (2009)
[arXiv:0904.2631 [hep-ph]].


\bibitem{Kobayashi:2017dyu} 
 T.~Kobayashi and S.~Nagamoto,
 Phys.\ Rev.\ D {\bf 96}, no. 9, 096011 (2017)
 [arXiv:1709.09784 [hep-th]].
 
\bibitem{Kobayashi:2018rad} 
 T.~Kobayashi, S.~Nagamoto, S.~Takada, S.~Tamba and T.~H.~Tatsuishi,
 Phys.\ Rev.\ D {\bf 97}, no. 11, 116002 (2018)
 [arXiv:1804.06644 [hep-th]].
 


\bibitem{Kobayashi:2018bff}
T.~Kobayashi and S.~Tamba,
Phys.\ Rev.\ D {\bf 99} (2019) no.4, 046001
[arXiv:1811.11384 [hep-th]].


\bibitem{Ohki:2020bpo}
H.~Ohki, S.~Uemura and R.~Watanabe,
[arXiv:2003.04174 [hep-th]].


\bibitem{Lauer:1989ax} 
  J.~Lauer, J.~Mas and H.~P.~Nilles,
  Phys.\ Lett.\ B {\bf 226}, 251 (1989);
%
  Nucl.\ Phys.\ B {\bf 351}, 353 (1991).
  

\bibitem{Lerche:1989cs} 
  W.~Lerche, D.~Lust and N.~P.~Warner,
  Phys.\ Lett.\ B {\bf 231}, 417 (1989).


\bibitem{Ferrara:1989qb} 
  S.~Ferrara, .D.~Lust and S.~Theisen,
  Phys.\ Lett.\ B {\bf 233}, 147 (1989).



\bibitem{Baur:2019kwi} 
 A.~Baur, H.~P.~Nilles, A.~Trautner and P.~K.~S.~Vaudrevange,
 Phys.\ Lett.\ B {\bf 795}, 7 (2019)
 [arXiv:1901.03251 [hep-th]]; 
 %
 arXiv:1908.00805 [hep-th].

\bibitem{Nilles:2020nnc}
H.~P.~Nilles, S.~Ramos-Sánchez and P.~K.~Vaudrevange,
JHEP \textbf{02}, 045 (2020)
[arXiv:2001.01736 [hep-ph]];
%
[arXiv:2004.05200 [hep-ph]].


\bibitem{Kobayashi:2016ovu} 
  T.~Kobayashi, S.~Nagamoto and S.~Uemura,
  PTEP {\bf 2017}, no. 2, 023B02 (2017)
  [arXiv:1608.06129 [hep-th]].





\bibitem{Kariyazono:2019ehj} 
 Y.~Kariyazono, T.~Kobayashi, S.~Takada, S.~Tamba and H.~Uchida,
 Phys.\ Rev.\ D {\bf 100}, no. 4, 045014 (2019)
 [arXiv:1904.07546 [hep-th]].


\bibitem{Kobayashi:2020hoc}
T.~Kobayashi and H.~Otsuka,
Phys. Rev. D \textbf{101}, no.10, 106017 (2020)
[arXiv:2001.07972 [hep-th]].






\bibitem{deAdelhartToorop:2011re} 
R.~de Adelhart Toorop, F.~Feruglio and C.~Hagedorn,
Nucl.\ Phys.\ B {\bf 858}, 437 (2012)
[arXiv:1112.1340 [hep-ph]].






\bibitem{Feruglio:2017spp} 
  F.~Feruglio,
  arXiv:1706.08749 [hep-ph];
  J.~C.~Criado and F.~Feruglio,
  SciPost Phys.\  {\bf 5}, no. 5, 042 (2018)
  [arXiv:1807.01125 [hep-ph]];
%
  J.~C.~Criado, F.~Feruglio, F.~Feruglio and S.~J.~D.~King,
  arXiv:1908.11867 [hep-ph].

\bibitem{Kobayashi:2018vbk} 
  T.~Kobayashi, K.~Tanaka and T.~H.~Tatsuishi,
  Phys.\ Rev.\ D {\bf 98}, no. 1, 016004 (2018)
  [arXiv:1803.10391 [hep-ph]].

\bibitem{Penedo:2018nmg} 
  J.~T.~Penedo and S.~T.~Petcov,
  Nucl.\ Phys.\ B {\bf 939}, 292 (2019)
  [arXiv:1806.11040 [hep-ph]];
  P.~P.~Novichkov, J.~T.~Penedo, S.~T.~Petcov and A.~V.~Titov,
  JHEP {\bf 1904}, 005 (2019)
  [arXiv:1811.04933 [hep-ph]];
%
  P.~P.~Novichkov, J.~T.~Penedo, S.~T.~Petcov and A.~V.~Titov,
  JHEP {\bf 1904}, 174 (2019)
  [arXiv:1812.02158 [hep-ph]];
  P.~P.~Novichkov, S.~T.~Petcov and M.~Tanimoto,
  Phys.\ Lett.\ B {\bf 793}, 247 (2019)
  [arXiv:1812.11289 [hep-ph]].





\bibitem{Kobayashi:2018scp} 
  T.~Kobayashi, N.~Omoto, Y.~Shimizu, K.~Takagi, M.~Tanimoto and T.~H.~Tatsuishi,
  JHEP {\bf 1811}, 196 (2018)
  [arXiv:1808.03012 [hep-ph]];
%
  T.~Kobayashi, Y.~Shimizu, K.~Takagi, M.~Tanimoto and T.~H.~Tatsuishi,
  arXiv:1906.10341 [hep-ph];
%
  arXiv:1907.09141 [hep-ph];
  T.~Kobayashi, Y.~Shimizu, K.~Takagi, M.~Tanimoto and T.~H.~Tatsuishi,
  Phys.\ Rev.\ D {\bf 100}, no. 11, 115045 (2019)
  [arXiv:1909.05139 [hep-ph]].
%
  T.~Kobayashi, Y.~Shimizu, K.~Takagi, M.~Tanimoto, T.~H.~Tatsuishi and H.~Uchida,
  Phys.\ Lett.\ B {\bf 794}, 114 (2019)
  [arXiv:1812.11072 [hep-ph]];
%
  arXiv:1910.11553 [hep-ph].
%

\bibitem{deAnda:2018ecu} 
  F.~J.~de Anda, S.~F.~King and E.~Perdomo,
  arXiv:1812.05620 [hep-ph];
  I.~De Medeiros Varzielas, S.~F.~King and Y.~L.~Zhou,
  arXiv:1906.02208 [hep-ph];
  S.~F.~King and Y.~L.~Zhou,
  Phys.\ Rev.\ D {\bf 101}, no. 1, 015001 (2020)
  [arXiv:1908.02770 [hep-ph]];
S.~King, J.D. and S.~F.~King,
[arXiv:2002.00969 [hep-ph]].


\bibitem{Okada:2018yrn} 
  H.~Okada and M.~Tanimoto,
  Phys.\ Lett.\ B {\bf 791}, 54 (2019)
 [arXiv:1812.09677 [hep-ph]];
  arXiv:1905.13421 [hep-ph];
%
[arXiv:2005.00775 [hep-ph]].

%

%
\bibitem{Ding:2019xna} 
  G.~J.~Ding, S.~F.~King and X.~G.~Liu,
  Phys.\ Rev.\ D {\bf 100}, no. 11, 115005 (2019)
  [arXiv:1903.12588 [hep-ph]];
  JHEP {\bf 1909}, 074 (2019)
  [arXiv:1907.11714 [hep-ph]];
  G.~J.~Ding, S.~F.~King, X.~G.~Liu and J.~N.~Lu,
  JHEP {\bf 1912}, 030 (2019)
  [arXiv:1910.03460 [hep-ph]];
%
[arXiv:2004.12662 [hep-ph]].

\bibitem{Nomura:2019jxj} 
  T.~Nomura and H.~Okada,
  Phys.\ Lett.\ B {\bf 797}, 134799 (2019)
  [arXiv:1904.03937 [hep-ph]];
  arXiv:1906.03927 [hep-ph];
  H.~Okada and Y.~Orikasa,
  Phys.\ Rev.\ D {\bf 100}, no. 11, 115037 (2019)
  [arXiv:1907.04716 [hep-ph]];
  arXiv:1907.13520 [hep-ph];
  arXiv:1908.08409 [hep-ph];
%
  T.~Nomura, H.~Okada and O.~Popov,
  arXiv:1908.07457 [hep-ph];
  T.~Nomura, H.~Okada and S.~Patra,
  arXiv:1912.00379 [hep-ph];
H.~Okada and Y.~Shoji,
[arXiv:2003.13219 [hep-ph]].

\bibitem{Novichkov:2019sqv} 
  P.~P.~Novichkov, J.~T.~Penedo, S.~T.~Petcov and A.~V.~Titov,
  JHEP {\bf 1907}, 165 (2019)
  [arXiv:1905.11970 [hep-ph]].
%



\bibitem{Liu:2019khw} 
  X.~G.~Liu and G.~J.~Ding,
  JHEP {\bf 1908}, 134 (2019)
  [arXiv:1907.01488 [hep-ph]].
%


%
%
\bibitem{Asaka:2019vev} 
  T.~Asaka, Y.~Heo, T.~H.~Tatsuishi and T.~Yoshida,
  arXiv:1909.06520 [hep-ph].
%

\bibitem{Zhang:2019ngf} 
  D.~Zhang,
  arXiv:1910.07869 [hep-ph].

\bibitem{Wang:2019ovr} 
  X.~Wang and S.~Zhou,
  arXiv:1910.09473 [hep-ph].
%


%
\bibitem{Kobayashi:2019gtp} 
  T.~Kobayashi, T.~Nomura and T.~Shimomura,
  arXiv:1912.00637 [hep-ph].

\bibitem{Lu:2019vgm} 
  J.~N.~Lu, X.~G.~Liu and G.~J.~Ding,
  arXiv:1912.07573 [hep-ph].

\bibitem{Wang:2019xbo} 
  X.~Wang,
  arXiv:1912.13284 [hep-ph].


%
\bibitem{Abbas:2020qzc}
M.~Abbas,
[arXiv:2002.01929 [hep-ph]].
%

%
%






	\bibitem{Gunning:1962}
 	R. C. Gunning,
 	\textit{Lectures on Modular Forms}
 	(Princeton University Press, Princeton, NJ, 1962).


\bibitem{Schoeneberg:1974}
 B.~Schoeneberg,
  \textit{Elliptic Modular Functions}
  (Springer-Verlag, 1974)

\bibitem{Koblitz:1984}
 N.~Koblitz,
  \textit{Introduction to Elliptic Curves and Modular Forms}
  (Springer-Verlag, 1984)


\bibitem{Bruinier:2008}
J.H.~Bruinier, G.V.D.~Geer, G.~Harder, and D.~Zagier,
\textit{The 1-2-3 of Modular Forms}
(Springer, 2008)





\bibitem{Cremades:2004wa}
D.~Cremades, L.~Ibanez and F.~Marchesano,
JHEP \textbf{05} (2004), 079
[arXiv:hep-th/0404229 [hep-th]].

\bibitem{Hamada:2012wj}
Y.~Hamada and T.~Kobayashi,
Prog.\ Theor.\ Phys.\  \textbf{128} (2012), 903-923
[arXiv:1207.6867 [hep-th]].



\bibitem{shimura}
G.~Shimura,
Annals of Mathematics, 97(3), second series, 440 (1973)


\bibitem{Duncan:2018wbw}
J.~F.~Duncan and D.~A.~Mcgady,
[arXiv:1806.09875 [math.NT]].

\bibitem{Kubota}
T.~Kubota,
Kiyokuniya Book Store, 1969

\bibitem{Budden}
M.~Budden and G.~Goehle,
J.~Lie Theory 27 (2017), 657-670










\bibitem{Abe:2008fi}
H.~Abe, T.~Kobayashi and H.~Ohki,
JHEP \textbf{09} (2008), 043
[arXiv:0806.4748 [hep-th]].

\bibitem{Abe:2013bca} 
  T.~H.~Abe, Y.~Fujimoto, T.~Kobayashi, T.~Miura, K.~Nishiwaki and M.~Sakamoto,
  JHEP {\bf 1401}, 065 (2014)
  [arXiv:1309.4925 [hep-th]].
  
\bibitem{Abe:2014noa} 
  T.~h.~Abe, Y.~Fujimoto, T.~Kobayashi, T.~Miura, K.~Nishiwaki and M.~Sakamoto,
  Nucl.\ Phys.\ B {\bf 890}, 442 (2014)
  [arXiv:1409.5421 [hep-th]].
  




\bibitem{Fujimoto:2013xha}
Y.~Fujimoto, T.~Kobayashi, T.~Miura, K.~Nishiwaki and M.~Sakamoto,
Phys.\ Rev.\ D \textbf{87} (2013) no.8, 086001
[arXiv:1302.5768 [hep-th]].








\bibitem{Abe:2008sx}
H.~Abe, K.~S.~Choi, T.~Kobayashi and H.~Ohki,
Nucl. Phys. B \textbf{814}, 265-292 (2009)
[arXiv:0812.3534 [hep-th]].

\bibitem{Abe:2015yva}
T.~h.~Abe, Y.~Fujimoto, T.~Kobayashi, T.~Miura, K.~Nishiwaki, M.~Sakamoto and Y.~Tatsuta,
Nucl. Phys. B \textbf{894}, 374-406 (2015)
[arXiv:1501.02787 [hep-ph]].





\bibitem{Kobayashi:2020uaj}
T.~Kobayashi and H.~Otsuka,
[arXiv:2004.04518 [hep-th]].



\end{thebibliography}
\end{document}